%% file: v18.tex
\documentclass[12pt, onecolumn, draftclsnofoot]{IEEEtran}%

\input{preamble}

\usepackage{cite}
\usepackage{tikz}
\usepackage{pgfplots}
\pgfplotsset{compat=1.5}
\usepackage{breqn}
\usepackage{subfigure}
\providecommand{\U}[1]{\protect\rule{.1in}{.1in}}

\theoremstyle{definition}
\newtheorem{defn}{Definition}

\numberwithin{Assumption}{section}

\begin{document}
	\title {Network Topology Inference Using Information Cascades with Limited Statistical Knowledge}
	\author{Feng~Ji, Wenchang~Tang,  Wee~Peng~Tay,~\IEEEmembership{Senior Member,~IEEE}, and Edwin~K.~P.~Chong~\IEEEmembership{Fellow,~IEEE}%
			\thanks{This research is supported by the Singapore Ministry of Education Academic Research Fund Tier 1 grant 2017-T1-001-059 (RG20/17) for the first three authors; and the National Science Foundation grant CCF-1422658 for the last author.}%
		\thanks{F. Ji and W. Tang contributed equally to this work. F.~Ji, W.~Tang and W.~P.~Tay are with the School of Electrical and Electronic Engineering, Nanyang Technological University, 639798, Singapore (e-mail: jifeng@ntu.edu.sg, E150012@e.ntu.edu.sg, wptay@ntu.edu.sg). Edwin~K.~P.~Chong is with  Dept. of Electrical and Computer Engineering, Colorado State University, USA (e-mail: Edwin.Chong@ColoState.Edu).}
	}
	
	\maketitle
	
	\begin{abstract}
		We study the problem of inferring network topology from information cascades, in which the amount of time taken for information to diffuse across an edge in the network follows an unknown distribution. Unlike previous studies, which assume knowledge of these distributions, we only require that diffusion along different edges in the network be independent together with limited moment information (e.g., the means). We introduce the concept of a separating vertex set for a graph, which is a set of vertices in which for any two given distinct vertices of the graph, there exists a vertex whose distance to them are different. We show that a necessary condition for reconstructing a tree perfectly using distance information between pairs of vertices is given by the size of an observed separating vertex set. We then propose an algorithm to recover the tree structure using infection times, whose differences have means corresponding to the distance between two vertices. To improve the accuracy of our algorithm, we propose the concept of redundant vertices, which allows us to perform averaging to better estimate the distance between two vertices. Though the theory is developed mainly for tree networks, we demonstrate how the algorithm can be extended heuristically to general graphs. Simulations using synthetic and real networks,
		and experiments using real-world data suggest that our proposed algorithm performs better than some current state-of-the-art network reconstruction methods.
	\end{abstract}
	
	\begin{IEEEkeywords}
		Network topology inference, information cascades, information diffusion, graph theory
	\end{IEEEkeywords}
	
	\section{Introduction}\label{sec:Intro}
	
\blue{Complex networks inspired by empirical research on real-world networks in nature and human societies, have been studied extensively in recent years \cite{manoj2018, newman2003}. The theory of complex networks has found applications in diverse areas including theoretical physics, sociology and biology \cite{costa2011,Helbing2015,Gosak2018,Gui2013}.	There has been recent increased interest to study the spread or diffusion of information, influence, or infections across a network. For example, information dynamics and social learning have been investigated in \cite{Gui2013, SohTayQue:J13, Kempe2003, Java2006, Leskovec2007, Tay:J15, HoTayQue:J15, Gosak2018, Jalili2017,Helbing2015}. Gaining insights into such dynamics have many useful applications, including developing effective information dissemination strategies, rumor sanitization methods and viral marketing approaches. Disease spreading over complex networks has also been extensively studied \cite{manoj2018, newman2003, Helbing2015, Gui2013}. Learning how diseases spread allow us to develop better control mechanisms and regulatory policies. Finding the sources of an infection diffusing in a network \cite{Shah2011, Dong2013, LuoTayLeng13, LuoTay:C13, Lokho2013, LuoTayLen14, LuoTayLen:J16, JiTayVar:J17, TanTay:C17,TanJiTay:J18} has many practical applications, including helping government agencies to identify the culprits who started a malicious rumor, the failure points that lead to a cascading power grid blackout, and entry points of a virus or malware into a computer network. In all the before mentioned applications, we assume that the underlying network topology is known. However, in many practical applications, the network topology may not be known in advance and needs to be inferred from the infection times of the nodes.} 
	
In this paper, we consider the inference of a network topology based on knowledge of the time each vertex in the network receives a piece of information, when an information diffusion is initiated from a known source vertex. More precisely, the network is modeled by an undirected graph $G=(V,E)$, with $V$ the set of vertices and $E$ the set of edges. 
	A single source vertex initiates an information diffusion, where the information is spread from each ``infected'' vertex to its neighbors stochastically along the edges connecting them. The amount of time it takes for the information to reach each vertex of $G$ is observed. We call this the \emph{infection time} of the vertex, and a collection of infection times with their corresponding source vertex a \emph{cascade}. 
	We wish to use these information cascades to estimate the connections among vertices of the graph.
	
	The network topology inference problem using information cascades has been investigated under various assumptions. For example, \cite{Gomez2010, Gom2011} considered the inference of graphs with a continuous-time diffusion model. The diffusion along each edge is assumed to be exponential or satisfies a power law. In \cite{Abr2013}, the authors assumed that the network can be undirected, and, along each edge, information spreads in two steps: a Bernoulli selection step and an exponential transmission step. The papers \cite{MyeLes2010, Net2012} perform topology inference using likelihood maximization approaches. In all of these works, knowledge of the family of probability distributions that generates the information spreading along edges is assumed, and many also assumed the spreading along different edges to be identically distributed. Theoretical guarantees and recovery conditions are provided in the references \cite{Abr2013,Net2012,Daneshmand2016}. 
    \blue{Similar research work also includes link prediction in social networks such as Twitter \cite{Sanda2017, Jalili2017Link}. Link prediction refers to inferring the future relationships from nodes in the complex network based on the observed network structure and node attributes. Therefore, it assumes the knowledge of existing network topology and does not use cascades as observations.}
    
	In practice, there are situations where it is not easy to know what the spreading distribution looks like \emph{a priori}. For example, information diffusion in an online social network depends on a variety of factors \cite{Gui2013}. In some cases, \emph{moments} -- such as the average amount of time information takes to diffuse from a vertex to a neighbor -- can be estimated from historical data\cite{Gruhl2004, Centola1194}, \blue{data collected from a related network, or based on domain experts' opinions}. In this paper, we consider the case where the diffusion along different edges be independent, and the spreading distribution is unknown to an observer. However, certain moments of the distribution (instead of the full distribution) are known. Our assumption is complementary to those made in \cite{Gomez2010, Gom2011, Abr2013, Net2012} in the following sense: while these previous works assume that the spreading distribution belongs to a known family of distributions with unknown parameters or moments, our work assumes that some parameters of the spreading distribution are known but not the family it belongs to. Either set of assumptions has its own merits and maybe more suitable than the other under different applications. Our experiments in \cref{sec:sim} indicate that in some cases, the best topology inference strategy involves making an assumption about the family of distributions (which may be wrong and lead to the problem of distribution mismatch), estimating the moments required for our proposed approach, and then applying our proposed approach. 

In \cite{Dunips}, the authors proposed a kernel-based method that does not require knowledge of the spreading distributions. They kernelize the transmission functions over network edges and then infer them from the data, which requires a large amount of data for accurate inference. Another drawback is that the performance is sensitive to the choice of hyperparameters and kernels, which need to be adjusted for different networks and amount of observed data. Experiments in \cref{sec:sim} indicate that our proposed approach has a better performance.

Owing to recent advancements in the field of graph signal processing (see \cite{Shu13} for an overview), several network topology inference methods using graph signals have been proposed (for example, \cite{Dong15J,Kal16,Seg16J}). These approaches are based on the idea that certain graph signals are closely related to eigenvectors of graph-shift operators of a graph (e.g., graph adjacency matrix and graph Laplacian). However, in our setting, the timestamps are related to the distance from vertices of a graph to a fixed source node. Such timestamps are usually not directly related to the above-mentioned shift-operators, and depend on the choice of the source node. Therefore, the graph signal processing approaches cannot be applied easily to our problem. 
	
Our objectives in this paper are twofold: to understand when a graph $G$ is perfectly reconstructable from only information cascades and moment information, and to develop low-complexity algorithms to infer graph topologies. The first objective is essentially hopeless for general graphs given the limited amount of prior information we assume (note that \cite{Net2012,Daneshmand2016} assume that the spreading distribution is known). Therefore, we consider only the case where $G$ is a tree in the first objective. 
Our main contributions are as follows:
\begin{enumerate}[(i)]
	\item In the case where the network is a tree and the mean propagation time across every edge is the same, we derive a necessary condition for perfect tree reconstruction based only on distance information (which is proportional to the expected infection times). We introduce the concept of a \emph{separating vertex set}, show that if this set is sufficiently large, we can reconstruct the tree using the distances of vertices from those in the separating vertex set.
	\item We develop an iterative tree inference algorithm that makes use of information cascades and the average mean propagation time along each edge. We introduce	the concept of redundant vertices to reduce the estimation variance. Simulations suggest that our method outperforms the algorithm proposed in \cite{Abr2013}.
	\item Under some technical conditions, we show that it is possible to use a higher order moment of the propagation time from a vertex $u$ to another vertex $v$ to determine if $(u,v)\in E$. We then develop a heuristic general graph inference algorithm by extending our tree reconstruction method to make use of the higher order moments of the propagation time along each edge. Here, we assume that an estimate of the average node degree is available. Simulations suggest that our proposed approach has relatively low time complexity and good edge recovery rate compared to current state-of-the-art methods in \cite{Dunips,Gom2011}. 
\end{enumerate}

The rest of this paper is organized as follows. In \cref{sec:model}, we introduce our graph model and assumptions. In \cref{sec:sep}, we introduce the notion of a separating vertex set and derive a necessary condition for perfect tree reconstruction. In \cref{sec:re}, we introduce the notion of redundant vertices and use that to develop an iterative tree inference algorithm in \cref{sec:alg}. We further extend the algorithm to general graphs in \cref{sec:graph}. Theoretical results supporting the extension are also discussed in \cref{sec:graph}. We present simulation results and experiments on real data in \cref{sec:sim} and conclude in \cref{sec:con}. 
	
	\section{Graph model} \label{sec:model}	
	Let $G=(V,E)$ be a connected and unweighted simple graph (i.e., undirected graph containing no graph loops or multiple edges), with $V$ the set of vertices and $E$ the set of edges. Let $d(\cdot,\cdot)$ be the length of a shortest path between nodes $u$ and $v$. For each vertex $v$, let $d_v(u)=d(v,u)$ denote the distance function (from any other node $u\in G$) to $v$. We use $[u,v]$ to denote any shortest path between $u$ and $v$, and $(u,v)$ the path with end vertices $u,v$ excluded. A path $P$ between $u$ and $v$ is called \emph{simple} if $P$ does not cross itself; thus, $[u,v]$ is always a simple path. 
	
	A \emph{cascade} consists of a single source vertex $u$ that initiates an information diffusion together with the times $\{T_u(v) : v\in V\}$, where $T_u(v)$ is the amount of time it takes information to propagate from vertex $u$ to vertex $v$. We assume that the information diffusion along different edges are independent, and the average of some moments of the propagation time along each edge are known \emph{a priori}. 
For each edge $(u,v)\in E$ and $k\geq 1$, let the $k$-th moment of the information propagation time along $(u,v)$ be $\mu\tc{k}_{u,v}$. We require to know $\mu\tc{k}$, the average of $\mu\tc{k}_{u,v}$ over the edges $(u,v)\in E$, for $k=1$ if $G$ is a tree, and for $k=1,2$ if $G$ is a general graph.
This assumption is valid in the following application scenarios:
	\begin{enumerate}[(a)]
		\item\label{it:a} Historical data is available for us to compute the empirical moments for information propagation between vertices in a network. Such data may be available only for specific vertices in the network whose neighbors are known, or for vertices in a different network that shares similar characteristics as the network of interest. In this case, we simply take $\mu\tc{k}$ to be the historical empirical moment. In simulations in \cref{subsec:Heterogeneous} using the mean and variance of the propagation time, we demonstrate that this approach can still produce reasonable results even if the actual mean and variance for each edge is unknown and heterogeneous over the edges. A heuristic basis for this observation is as follows: Consider the case where $G$ is a tree, and suppose that the mean propagation time $\mu\tc{1}_{u,v}$ along any edge $(u,v)\in E$ is independently and identically distributed (i.i.d.) with mean $\mu$. For any pair of vertices $u$ and $v$, there is a unique simple path connecting them. Suppose it consists of $d$ edges $e_1,\ldots, e_d$. Let $t_{e_i}$ denote the time it takes to pass the infection along $e_i$. Considering the cascade initiated from $u$, we have
		\begin{align}
		\E[T_u(v)]{(\mu\tc{1}_{e_i})_{i=1}^d} &= \sum_{1\leq i\leq d}\E[t_{e_i}]{\mu\tc{1}_{e_i}}\nonumber\\ 
		&= \sum_{1\leq i\leq d}\mu\tc{1}_{e_i}.\label{Tu_mean}
		\end{align}
		Applying the strong law of large numbers \cite{Dur:95} to the sum \eqref{Tu_mean}, $\E[T_u(v)]{(\mu\tc{1}_{e_i})_{i=1}^d}/d \to \mu$ almost surely as $d\to\infty$. If we adopt the stochastic approximation that the observed $T_u(v) \approx \allowbreak \E[T_u(v)]{(\mu\tc{1}_{e_i})_{i=1}^d} \allowbreak\approx \mu d$ for vertices far apart, then using the common mean $\mu$ allows us to estimate the distance $d$ between the vertices. We provide an example in \cref{subsec:real_world}, where we test our algorithm on a real-word dataset with the empirical mean and second order moment of the information propagation computed from historical data.
		
		\item We estimate the propagation time moments by first running a separate statistical estimation procedure like the NetRate algorithm proposed in \cite{Gom2011}. We then use the estimated propagation time moments for our proposed approach. In simulations in \cref{subsec:Heterogeneous}, we first run NetRate and then fuse our results with those produced by NetRate. We observe that this yields significant improvements over NetRate, even though the additional computation overhead incurred by running our method is negligible compared to NetRate.
	\end{enumerate}

	We assume that each cascade persists long enough to infect all the vertices, and cascades are initiated from vertices in a subset $V_c \subset V$. The subset $V_c$ is called the \emph{source set}. The timestamp information $\{T_u(v)\in \Real: u \in V_c, v\in V\}$ is recorded. Our goal is to infer the adjacency matrix $A_G$ of $G$ from the timestamp information. The discussion in \ref{it:a} above suggests that we may study the distance functions associated with vertices of the graph when the mean propagation times along edges are the same. In the sequel, we first develop the underlying theory and procedure for the case where only distance information is available, and then extend our procedure to the general case where the mean propagation times may be heterogeneous.
	
\section{Separating vertex set and reconstruction accuracy for trees}\label{sec:sep}
	
\begin{figure}[!t] 
		\centering
		\includegraphics[scale=0.7]{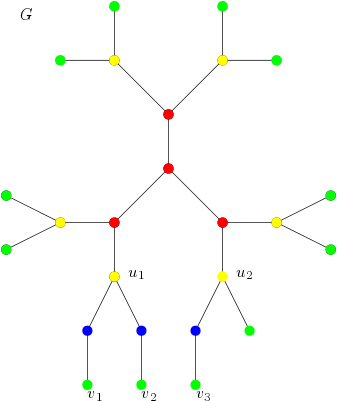}
		\caption{\blue{This figure illustrates concepts in \cref{defn:tso} and \cref{defb:abw}.} In $G$, the boundary $\partial G$ consists of the green vertices. The branched vertices $B_G$ are the red and yellow nodes, and $\partial B_G$ are the yellow nodes. The blue nodes are ordinary. } \label{fig:1}
	\end{figure}

In this section, we consider the case where the graph $G$ is a tree and the mean propagation time across every edge is the same. Therefore, the information $\{\E[T_u(v)]\in \Real: u \in V_c, v\in V\}$ is equivalent to $\Delta=\{d_u(v)\in \Real: u \in V_c, v\in V\}$. We develop conditions for perfect reconstruction of $G$ based on $\Delta$. From the strong law of large numbers, our results can then be said to hold with probability approaching one when the number of cascades becomes large. We start off with the following notion of ``distance" associated with any subset of vertices of the graph. For later use, we introduce the following definitions (see \cref{fig:1} for an example). 
	
	\begin{defn} \label{defn:tso}
		The set of \emph{leaf} or \emph{boundary} vertices, denoted by $\partial G$, are vertices of degree $1$. The set of \emph{branched} vertices, denoted by $B_G$, are vertices of degree at least $3$. The remaining vertices in $V\backslash (\partial G \cup B_G)$ are of degree $2$ and are called \emph{ordinary} vertices.
	\end{defn}
	
	\begin{defn} \label{defb:abw}
		We take $\partial B_G\subset B_G$ to be the subset of vertices having a simple path to a leaf without passing through other vertices of $B_G$: 
		\[
		\partial B_G = \{v \in B_G \mid [u,v] \cap B_G = \{v\} \text{ for some }u \in \partial G\}.
		\]
	\end{defn}
	
	Although $\partial B_G$ is defined for any graph, it is particularly useful when considering tree networks. 	
	
	\begin{defn}\label{def:conv}
		Given a subset of vertices $V'\subset V$, the \emph{convex hull} $\conv(V')$ of $V'$ in $V$ is the union of all simple paths connecting any pair of distinct vertices of $V'$. 
	\end{defn}
	
	\begin{defn} \label{defn:lvb}
		Let $V'=\{v_1,\ldots, v_l\}\subset V$. For any two nodes $u$ and $v$, define their \emph{relative distance} with respect to (w.r.t.) $V'$ as $$d_{V'}(u,v) = \sup_{v_i\in V'}{|d_{v_i}(u)-d_{v_i}(v)|}.$$ 
	\end{defn}
	
	By the triangle inequality associated with the usual absolute value, it is easy to verify that $d_{V'}(\cdot,\cdot)$ defines a pseudometric (which means that $d_{V'}(u,v)$ can be $0$ for $u\neq v$) on $V$.
	
\begin{figure}[!t] 
		\centering
		\includegraphics[scale=0.55]{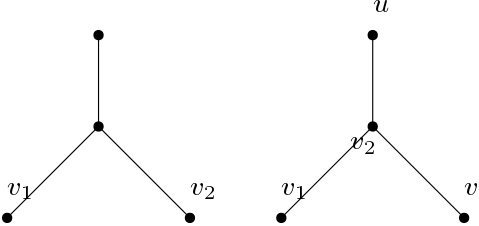}
		\caption{\blue{This figure illustrates the concept of a separating vertex set in \cref{defn:separate}.} The tree on the left is separated by $v_1$ and $v_2$. The tree on the right is not, because vertices $v$ and $u$ have the same distances to both $v_1$ and $v_2$.} \label{fig:2}
	\end{figure}

	In this section, using $d_{V'}(\cdot,\cdot)$, we aim to develop a necessary condition under which a tree $G$ can be reconstructed uniquely. To this end, we introduce the concept of a separating vertex set as follows (see \cref{fig:2}).
	
	\begin{defn} \label{defn:separate}
		A set $V'=\{v_i, 1\leq i\leq l\}$ \emph{separates} $G$ if for any distinct vertices $u,v\in V$, there exists $v_i \in V'$ such that $d_{v_i}(u)\neq d_{v_i}(v)$. We say that $V'$ is a \emph{separating vertex set}.
	\end{defn} 
	
	In the related source localization problem, a similar concept has been proposed. For example, in \cite{spinelli2017,chen2014Appro}, the authors define a Double Resolving Set (DRS). A DRS is a subset $Z\subseteq V$ such that for every $u,v\in V$ there exist $z_1,z_2\in Z$ such that $d_{z_1}(u)-d_{z_1}(v)\neq d_{z_2}(u)-d_{z_2}(v)$. This is not an equivalent definition to our separating vertex set, as there exists a separating vertex set that is not a DRS and vice versa.
	Intuitively, to infer a graph is equivalent to knowing the metric on the graph. We hope to infer the graph structure by using the distances to only the nodes in $V'$. The ``separating condition'' makes sure that $V'$ is typical enough. In the following, we demonstrate how to approximate $G$ using a subset $V'$ and the pseudometric $d_{V'}(\cdot,\cdot)$, and how the notion of ``separating vertex set'' are used. 
	
	\begin{defn}\label{def:reconstruction}
		Let $V' = \{v_1,\ldots,v_l\}$. We say that a graph $G'=(V,E')$ is \emph{reconstructed from} $V'$ if any two vertices $u,v \in V$ are connected by an edge in $E'$ if and only if $d_{V'}(u,v)\leq 1$.
	\end{defn}
	
	\begin{figure}[!htb] 
		\centering
		\includegraphics[scale=0.55]{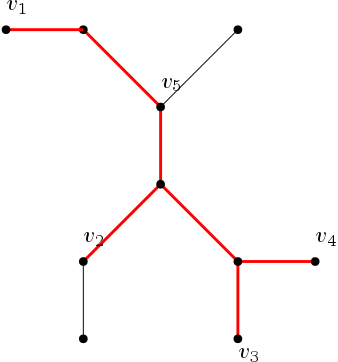}
		\caption{This figure illustrates the condition of \cref{thm:lvca}\ref{it:b}. Suppose $V'=\{v_1,\ldots,v_5\}$. The red edges form $\conv(V')$. Clearly, $\partial B_G' = \{v_5\}$ and $\partial \conv(V') = \{v_1,\ldots, v_4\}$. Therefore, because $V'$ separates $G$, the conditions of \cref{thm:lvca}\ref{it:b} are satisfied.} \label{fig:11}
	\end{figure}
	
	\begin{Theorem} \label{thm:lvca}
		Let $V' = \{v_1,\ldots,v_l\}$, and $G'=(V,E')$ be reconstructed from $V'$. Then the following holds true:
		\begin{enumerate} [(a)]
			\item $E \subset E'$.
		\item \label{it:sgia} Suppose that $G$ is a tree. Then $V'$ separates its convex hull $\conv(V')$.
			\item \label{it:b}Suppose that $G$ is a tree. Let $\partial B_G'$ contain vertices of $\partial B_G$ such that for each $v\in \partial B_G'$, some neighbors of $v$ are not in $\conv(V')$ (cf.\ \cref{fig:11}). Moreover, assume that $d(u,v)>1$ for $u,v \in \partial B_G'\cup \partial \conv(V')$. If $V'$ separates $G$, then $E' = E$, and hence $G'=G$.
			\item \label{it:c}In the converse direction, suppose that $G$ is a general graph that does not contain any triangle (three pairwise connected vertices). If $E'=E$, then $V'$ separates $G$.
		\end{enumerate}
	\end{Theorem}
	\begin{IEEEproof}
	See \cref{proof:thm:lvca}.
	\end{IEEEproof}
    
\Cref{thm:lvca} tells us when we can reconstruct $G$ perfectly by means of \cref{def:reconstruction} in terms of a ``separating vertex set". On the other hand, the size of a separating can be estimated using \cref{thm:lvca}. 
	
We introduce the following quantifier to evaluate the effectiveness of a given $V'$ from which $G$ is reconstructed.
	
	\begin{defn}\label{def:perfect_reconstruction}
		Let $V' \subset V$ and $G'=(V,E')$ be reconstructed from $V'$. We say that a subset of vertices $V_0\subset V$ is \emph{perfectly reconstructed} from $V'$ if the subgraphs spanned by $V_0$ in $G'=(V,E')$ and $G=(V,E)$ are the same. The \emph{reconstruction accuracy} of $V'$ is defined as 
		\begin{align}
		\psi(V') = \sup\left\{\frac{|V_0|}{|V|} : V_0 \text{ perfectly reconstructed from } V'\right\}.
		\end{align}
	\end{defn}
	
	We have the following observations regarding $\psi$\ based on the definition and results obtained so far. 
	
	\begin{Corollary} \label{coro:teg}
		Let $V' \subset V$.
		\begin{enumerate} [(a)]
			\item \label{it:tega}The entire graph $G=(V,E)$ is perfectly reconstructed from $V'$ if and only if $\psi(V')=1$.
			\item If $G$ is a tree and $\psi(V')=1$, then $V'$ is a separating vertex set.
			\item \label{it:tegc}If $G$ is a tree and $G = \conv(V')$, then $\psi(V')=1$.
		\end{enumerate}
	\end{Corollary}
	
	\begin{IEEEproof}\ 
		\begin{enumerate} [(a)]
		\item Follows immediately from the definition.
		
		\item Follows from \cref{thm:lvca}\ref{it:c}.
		
		\item The result follows from \ref{it:sgia} and noting that $\partial B_G' = \emptyset$ in \cref{thm:lvca}\ref{it:b}.
		\end{enumerate}
	\end{IEEEproof}
	
	\begin{Remark} \label{rmk:iwu}
		The condition $d_{V'}(u,v)\leq 1$ in \cref{def:reconstruction} is equivalent to $|d_{v_i}(u)-d_{v_i}(v)|\leq 1$ for all $v_i\in V'$. On the other hand, if $G$ is a tree, it is clear that for two vertices $u,v$ connected by an edge, $d(u,u')\neq d(v,u')$ for any $u' \in G$. Therefore, we can change the condition $d_{V'}(u,v)\leq 1$ in \cref{def:reconstruction} to $d_{V'}(u,v)=1$ if $G$ is a tree.
	\end{Remark}

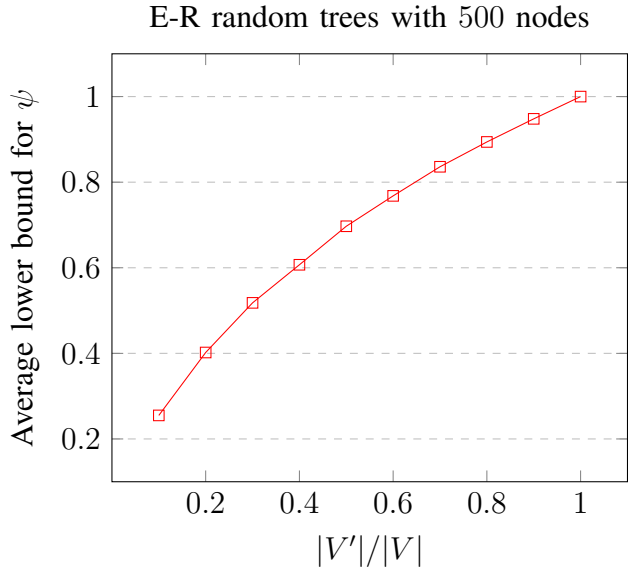
\begin{figure}[!htb]
\centering
\begin{tikzpicture}[scale = 1][baseline]
\begin{axis}[
title={E-R random trees with $500$ nodes},
xlabel={$\lvert V'\rvert/\lvert V \rvert$}, 
ylabel={Average lower bound for $\psi$},
xmin=0, xmax=1.1,
ymin=0.1, ymax=1.1,
xtick={0.2,0.4,0.6,0.8,1.0},
ytick={0.2,0.4,0.6,0.8,1.0},
legend pos=north west,
ymajorgrids=true,
grid style=dashed,
]
\addplot
[
color = red,
mark = square,
]
coordinates{
	(0.1,0.255)(0.2,0.402)(0.3,0.518)(0.4,0.607)(0.5,0.697)(0.6,0.768)(0.7,0.836)(0.8,0.894)(0.9,0.948)(1,1) 
};
\end{axis}
\end{tikzpicture}
\caption{The figure shows the average lower bound $|\conv(V')|/|V|$ of $\psi(V')$ versus $|V'|/|V|$.} \label{fig:rt}
\end{figure}

As an example, we perform a numerical experiment by constructing $200$ random trees with $500$ vertices each. Starting from one vertex, we add a new vertex in every step and attach it to one of the existing vertices randomly to obtain a non-scale-free tree. We call \blue{this} the Erd\H{o}s-R\'{e}nyi (E-R) tree.
We found that on average if the size of $V'$ is less than $0.16|V|$, then the tree cannot be determined uniquely. This suggests that perfect reconstruction of the entire network is in general difficult if insufficient information is available. For the interested reader, further insights are provided in the supplementary discussions in \cref{sup}.
	
For any given subset $V'$ of a tree, from \cref{coro:teg}\ref{it:tegc}, we have $\psi(V') \geq |\conv(V')|/|V|$.  Using the E-R random trees constructed in our experiment above, and choosing the subset $V'$ randomly, we plot $|\conv(V')|/|V|$ versus $|V'|/|V|$ in \cref{fig:rt}. In practice, we cannot determine a separating vertex set \emph{a priori}, and achieving $\psi(V')=1$ is usually impossible. However, from \cref{fig:rt}, we see that on average a set $V'$ with reasonable size allows perfect reconstruction of a large part of the network (e.g., observing $40\%$ of the network yields on average a $\psi(V') > 60\%$). In \cref{sec:alg}, we propose an algorithm that does not require a separating vertex set explicitly.

	\section{Redundant vertices}\label{sec:re}
	
	Our tree inference algorithm is based on the infection times of cascades, whose means are proportional to the distances between vertices. In practice, we face the following problems: (1) Owing to the stochastic nature of the diffusion process, the recorded infection times at various vertices are not exactly proportional to the distance to the source. (2) From \cref{sec:re}, in order to achieve a reasonable rate of recovery, we do not have to use all the nodes of $G$. In other words, there are ``redundant" nodes. In this section, we introduce and discuss the concept of redundant vertices before presenting a method to mitigate the aforementioned shortcomings in our tree inference algorithm in \cref{sec:alg}.

\begin{figure}[!t] 
	\centering
	\includegraphics[scale=0.5]{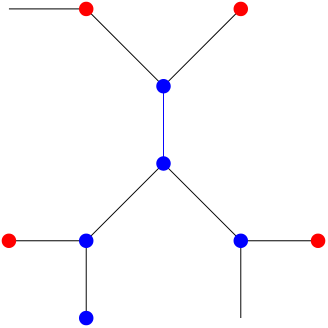}
	\caption{\blue{This figure illustrates the concept of a redundant vertex in \cref{defn:red}. In the tree,} $V'$ is the union of red and blue nodes. We may take the red nodes as a separating vertex set. On the other hand, all the blue nodes are redundant.} \label{fig:16}
\end{figure}

\begin{defn} \label{defn:red}
Let $V' =\{v_1,\ldots,v_l\}$ be a set of vertices. A vertex $v_i\in V'$ is called \emph{redundant} w.r.t.\ $V'$ if $E'$ reconstructed in the sense of \cref{def:reconstruction} using $V'$ and $V'\setminus \{v_i\}$ are the same (cf.\ \cref{fig:16}). Two vertices $v_1,v_2 \in V'$ are called \emph{mutually replaceable} if $E'$ reconstructed from $V'\setminus \{v_1\}$ and $V'\setminus \{v_2\}$ are the same. 
\end{defn}

Intuitively, when two vertices are mutually replaceable, they give the same amount of information in the reconstruction task. From the definition, we have the following simple observations. 
	
	\begin{Lemma} \label{lem:avvi}
		Let $V' =\{v_1,\ldots,v_l\} \subset V$.
		\begin{enumerate}[(a)]
			\item \label{it:avvi} A vertex $v_i \in V'$ is redundant if and only if for each pair $u,v$ such that $|d_{v_i}(u)-d_{v_i}(v)|>1,$ there is $v_j\in V'\setminus \{v_i\}$ such that $|d_{v_j}(u)-d_{v_j}(v)|>1$.
			\item If $v_i$ is redundant in $V'$, then $v_i$ is redundant in any set of vertices containing $V'$. 
			\item Two vertices $v_1, v_2 \in V'$ are mutually replaceable if and only if both are redundant w.r.t.\ $V'$.
		\end{enumerate}
	\end{Lemma}
	\begin{IEEEproof}
	See \cref{proof:sec:re}.
	\end{IEEEproof}
	
	\begin{defn}\label{def:subtree}
		Suppose that $G$ is a tree. For $u, v \in V$, remove the first edge in the path $[u,v]$, then we obtain two subtrees of $G$. We define $\calT_u^v$ as the subtree that contains $u$.
	\end{defn}	
	
	\begin{Proposition} \label{prop:sivi}
		Suppose that $G$ is a tree, and $V' =\{v_1,\ldots,v_l\} \subset V$. Let $v_i'$ be the vertex in $\conv(V'\setminus \{v_i\})$ closest to $v_i$ (i.e., $v_i'= \argmin_{v\in \conv(V'\setminus \{v_i\})} d(v,v_i)$). 
		\begin{enumerate}[(a)]
			\item If $v_i=v_i'$ (i.e., $v_i \in \conv(V'\setminus\{v_i\})$, then $v_i$ is redundant w.r.t.\ $V'$.
			\item Let $D_{v_i'}(r) = \{v\in V\mid d(v,v_i')\leq r\}$, and $\calT_{v_i'}^{v_i}$ be the subtree rooted at $v_i'$ pointing away from $v_i$. If $v_i\neq v_i'$, and the open path $P=(v_i,v_i')$ contains only ordinary vertices and $D_{v_i'}(2)\cap \calT_{v_i'}^{v_i} \subset \conv(V'\setminus\{v_i\})$, then $v_i$ is redundant w.r.t.\ $V'$.
		\end{enumerate}
	\end{Proposition}
	\begin{IEEEproof}
	See \cref{proof:sec:re}.
	\end{IEEEproof}
	
Using \cref{prop:sivi}, we obtain the following corollary.

\begin{Corollary} \label{coro:sgiat}
Suppose that $G$ is a tree. Let $\partial B_G'$ be any subset of $\partial B_G$ such that the distance between any two vertices of $\partial B_G'$ is at least $2$. Moreover, if each element of $\partial B_G'$ has at most two non-leaf neighbors, then any $V'$ contains a subset $V''$ of size at most $|\partial G| - |\partial B_G'|$ (which is a constant that depends only on $G$) such that all the vertices not in $V''$ are redundant w.r.t.\ $V'$.
\end{Corollary}
\begin{IEEEproof}
See \cref{proof:sec:re}.
\end{IEEEproof}
	
Based on this result, we run simulations on randomly generated trees with 500 nodes. The result show that on average, any $V'$ contains a subset $V''$ of size $32\%|V|$ such that the rest of nodes in $V'$ are redundant. 

	
	
\begin{figure}[!htb] 
			\centering
			\includegraphics[scale=0.5]{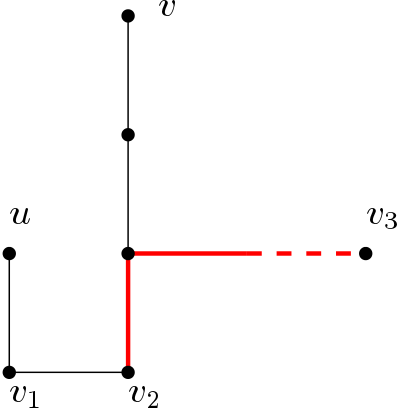}
			\caption{\blue{This figure corresponds to \cref{eg:asy}.} In the figure, $v_3$  tells that the nodes $u$ and $v$ cannot be connected by a direct edge; while $v_2$ does not. Therefore, we can conclude that $v_1$ is redundant if we have the additional node $v_3$ available.} \label{fig:15}
		\end{figure}

	\begin{Example} \label{eg:asy}
		A simple yet important example is when $V'$ contains two vertices $v_1,v_2$ connected by a direct edge as shown in \cref{fig:15}. Assume that $|V'|\geq 3$. Let $v_3$ be different from $v_1$ and $v_2$. Because $v_1$ and $v_2$ are connected by a direct edge, without loss of generality, we can assume that $v_2 \in \conv\{v_1,v_3\}=[v_1,v_3]$. By \cref{prop:sivi}(a), $v_2$ is redundant w.r.t.\ $V'$. In the following cases, we can also conclude that $v_1$ is redundant:
		\begin{enumerate}[(i)]
			\item Either (a) or (b) of \cref{prop:sivi} holds for $v_2$. Notice that $v_1'$ is either $v_1$ or $v_2$.
			\item \label{it:iwc} If we construct $E'$ using $|d_{v_i}(u)-d_{v_i}(v)|=1$, then $D_{v_i'}(2)$ in \cref{prop:sivi}(b) can be replaced by $D_{v_2}(1)$. See \cref{proof:sec:re} for a proof.
		\end{enumerate} 	
		If any of the above cases hold, and information diffusion happens deterministically, $v_1$ and $v_2$ are mutually replaceable w.r.t.\ $V'$ in inferring the structure of the graph. Simulation results show that if $|V'|\geq 0.3|V|$, then on average, more than $84\%$ of pairs of $V'$ connected by a direct edge are mutually replaceable. 
	\end{Example}

    The large amount of redundant vertices does not mean that cascades from these vertices are useless. In practice, the diffusion process is stochastic and we cannot guarantee that cascades are initiated from a fixed separating vertex set. If two nodes $v_1$ and $v_2$ are mutually replaceable, time information provided by $v_1$ (respectively, $v_2$) can be used to average out the noise in the time information at $v_2$ (respectively, $v_1$) to obtain a better estimate of the distance with a lower variance. In the next section, we make use of this idea to develop a tree inference algorithm. 
	
\section{Iterative Tree Inference Algorithm} \label{sec:alg}
	
Recall that for a cascade starting at vertex $u$, we use $T_u(v)$ to denote the first time that node $v$ receives the information of the cascade. Recall also that $V_c =\{v_1,\ldots,v_l\}$ is the set of vertices containing the sources of all the cascades. If there are several cascades with the same source, we average their infection times. The timestamp information is therefore $\{T_{v_i}(\cdot): v_i\in V_c\}$. From our discussion in the previous sections, in order to use infection-time information associated with cascades from various sources effectively, we propose the following general scheme if $G$ is a tree. More details are provided in the following discussion. 
	\begin{enumerate}[(1)]
		\item {\bf Selection step}: Given $V_c = \{v_1,\ldots,v_l\}$, find pairs of vertices $(v_i,v_j)$ having a ``high chance'' of being connected by a single edge.
		\item {\bf Transfer step}: For a pair of vertices $(v_i,v_j)$ connected by an edge, if $v_i, v_j \in V_c$, i.e., cascades at both $v_i$ and $v_j$ exist, we use the cascade at $v_i$ to construct a new cascade at $v_j$, and average the infection times with those from the existing cascade at $v_j$. This is where we use the concept of redundant vertices developed in \cref{sec:re} (cf.\ \cref{eg:asy}). 
		\item {\bf Reconstruction step}: Use the new timestamp information obtained in the previous steps to estimate $d_{v_i}(\cdot)$ for each $v_i\in V_c$, and reconstruct the graph $G$ based on \cref{def:reconstruction}.
	\end{enumerate}
	
	We first discuss the {\bf reconstruction step}. From \cref{thm:lvca} and \cref{rmk:iwu}, to determine if $u,v$ are connected by an edge, we want to compare $|d_{v_i}(u)-d_{v_i}(v)|$ with $1$ for each $v_i\in V_c$. This motivates us to introduce the following weight for each pair $u,v$: 
	\begin{align}\label{eq:edge-rec}
	W(u,v)\triangleq\frac{\sum_{v_i \in V_c}\left|  |T_{v_i}(u)-T_{v_i}(v)|/\mu\tc{1}-1 \right| }{|V_c|},
	\end{align}
	where we recall that $\mu\tc{1}$ is the average mean propagation time across each edge. Here, we take the average difference with $1$ instead of using $\sup$ as in \cref{defn:lvb}. This is because when the timestamps are stochastically generated, taking $\sup$ is sensitive to the noise inherent in the timestamps. 
	
If $W(u,v)$ is small, it suggests a higher chance that $u,v$ are connected by a direct edge. We call $W$ the \emph{weight matrix}, and its entries the \emph{weights}. If we want to infer the structure of a tree of size $n$, we can select $n-1$ (total number of edges) pairs of distinct vertices $(u,v)$ such that $W(u,v)=W(v,u)$ are the smallest $n-1$ weights. 

	Recall that we have assumed that each cascade persists long enough to infect all the vertices for theoretical convenience. In the algorithm application, this assumption can be relaxed. Given $u,v$ and $v_i\in V_c$, if $T_{v_i}(u)$ or $T_{v_i}(v)$ does not exist, we skip this cascade when computing $W(u,v)$. Therefore, in \eqref{eq:edge-rec} for each pair of $(u,v)$, we replace $V_c$ with $V_c(u,v)=\{v_i\in V_c: T_{v_i}(v)\neq\emptyset \text{ and } T_{v_i}(u)\neq \emptyset\}$. It is easy to see that the variance of $W(u,v)$ is greater with smaller $|V_c(u,v)|$, making it harder to judge whether an edge exists between $u$ and $v$. In applications, if $|V_c(u,v)|\ll |V_c|$ for some $(u,v)$, we recommend to ignore such $W(u,v)$.
	
Let the estimated graph be $G'=(V,E')$. We can directly use $E'$ for the {\bf selection step}. An alternative is to use the weight matrix $W$ for the job. More precisely, we can set a numerical condition $\calC$ based on $W$, and select a pair $(u,v)$ as long as $W(u,v)=W(v,u)$ satisfies the pre-set condition $\calC$. For example, for a fixed parameter $k$, we can choose $kn$ pairs of $(u,v)$ with the smallest $W(u,v)$ value. This reflects our belief that each selected $(u,v)$ has a ``high chance" of being connected by an edge. 
	
	For a selected pair of vertices $(v_i,v_j)$ with $v_i,v_j\in V_c$, and $u\in V$, we construct the times $T^{v_i}_{v_j}(\cdot)$ at $v_j$ using $T_{v_i}$ in the {\bf transfer step} as follows: for each $u\in V$,
	\begin{align}\label{cas-gen}
	T_{v_j}^{v_i}(u)&=\argmin_{x\in\left\lbrace T_{v_i}(u)+T_{v_i}(v_j),T_{v_i}(u)-T_{v_i}(v_j)\right\rbrace}|x-T_{v_j}(u)|. 
	\end{align}
	We now update $T_{v_j}(\cdot)$ by its average with $T^{v_i}_{v_j}(\cdot)$. Once $T_{v_j}(\cdot)$ are updated for every $v_j\in V_c$, we repeat the reconstruction step. 

\begin{figure}[!htb] 
		\centering
		\includegraphics[scale=1.3]{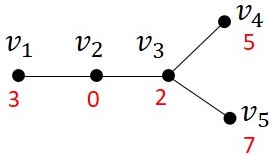}
		\caption{\blue{This figure gives an example to explain why we use \eqref{cas-gen} in the transfer step.} There are 5 vertices in the tree. We have a cascade initiated by $v_2$, with the red numbers being $\{T_{v_2}(\cdot)\}$. Now we want to construct $T_{v_3}^{v_2}(\cdot)$. For vertices $v_1$ and $v_2$, let $T_{v_3}^{v_2}(\cdot)=T_{v_2}(\cdot)+T_{v_2}(v_3)$. For vertices $v_3$, $v_4$ and $v_5$, let $T_{v_3}^{v_2}(\cdot)=T_{v_2}(\cdot)-T_{v_2}(v_3)$. Then for $T_{v_2}(\cdot)$ and $T_{v_3}^{v_2}(\cdot)$, the propagation times along all edges are all the same, but they are initiated by different sources. However, if we do not know the graph topology in advance, then for each vertex $v_j$, whether to choose $T_{v_2}(v_j)-T_{v_2}(v_3)$ or $T_{v_2}(v_j)+T_{v_2}(v_3)$ is unknown \emph{a priori}. We propose to compare with a cascade initiated by $v_3$.} \label{fig:transtep}
	\end{figure}

In \cref{cas-gen} of the transfer step, we consider a cascade $\{T_{v_i}(\cdot)\}$ from $v_i$ and attempt to reinterpret the time information as an observed cascade from $v_j$ instead. If the observed infection was initiated by $v_j$, the term $T_{v_i}(u)+T_{v_i}(v_j)$ in \cref{cas-gen} is the amount of time information takes to propagate from $v_j$ to $u$ if $v_i \in [v_j,u]$, while $T_{v_i}(u)-T_{v_i}(v_j)$ applies if $v_j \in [v_i,u]$. To determine which of these two cases is more likely, we compare them to the observed $T_{v_j}(u)$. An example is shown in \cref{fig:transtep} for illustration.

In the transfer step, note that no additional observations other than the given $\{T_{v_i}(\cdot): v_i\in V_c\}$ are used in our inference procedure. The transfer step is merely an averaging mechanism over the given timestamp information $\{T_{v_i}(\cdot): v_i\in V_c\}$ that allows us to average out the the noise in $T_{v_j}(\cdot)$ for some $v_j\in V_c$. The intuition is that if two vertices $v_i$ and $v_j$ are connected by an edge in a tree, then the cascades generated by them are mutually ``transferable''. We modify $T_{v_i}(\cdot)$ to obtain $T_{v_j}^{v_i}(\cdot)$ using \eqref{cas-gen} and regard $T_{v_j}^{v_i}(\cdot)$ as another cascade ``generated'' by $v_j$. An average with $T_{v_j}(\cdot)$ is then taken, thus reducing the randomness inherent in the timestamps.  Simulation results in \cref{Subsec:sim_tree} \cref{fig:six} show that the performance can be greatly improved if we perform additional transfer steps.

	We summarize the above discussion as our Iterative Tree Inference (ITI) algorithm in \cref{algo:tia}.
	
	\begin{algorithm}[!htb]
		\caption{Iterative tree inference (ITI) algorithm} \label{algo:tia}
		\begin{algorithmic}[1]		
			\STATE {\bf Input} $I:$ number of transfer steps, $m_s:$ number of edges selected for the selection step, $n:$ number of vertices, $V_c$: set of cascade sources, $\{T_{v_i}(\cdot): v_i\in V_c\}$: timestamp information, $\mu\tc{1}$: average mean propagation time.
			\STATE Initialize $\eta_{v}=0$ for each $v \in V_c$.
			\FOR {$s=1,\ldots,I$}
			\STATE Compute weight matrix $W$ in \eqref{eq:edge-rec} and select $m_s$ pairs $\left\lbrace (v_i,v_j)\right\rbrace $ having the $m_s$ smallest $W(v_i,v_j)$ weights.\\
			\FOR {Each pair of $(v_i,v_j)$ selected} 
			\STATE Construct $T_{v_i}^{v_j}(\cdot)$ and $T_{v_j}^{v_i}(\cdot)$ using \eqref{cas-gen}.\\
			\STATE Update $\eta_v=\eta_v+1$ for $v=v_i,v_j.$
			\STATE Update $T_{v_i}(\cdot) = (\eta_{v_i}T_{v_i}(\cdot) + T_{v_i}^{v_j}(\cdot))/(\eta_{v_i}+1)$ and $T_{v_j}(\cdot) = (\eta_{v_j}T_{v_j}(\cdot) + T_{v_j}^{v_i}(\cdot))/(\eta_{v_j}+1)$.
			\ENDFOR
			\ENDFOR
			\STATE Select $n-1$ edges $\left\lbrace (u,v)\right\rbrace$ corresponding to the $n-1$ smallest $W(u,v)$ weights.
			\STATE {\bf Output} $\left\lbrace (u,v)\right\rbrace$.
		\end{algorithmic}
	\end{algorithm}

	\section{The case of general graphs} \label{sec:graph}
	
	In this section, we consider general simple graphs. We first discuss theoretical results on information propagation on general graphs, which lend support to our heuristic extension of the ITI algorithm. We then propose a general graph inference algorithm.
	
	\subsection{Theoretical observations} \label{subsec:to}
	
	To facilitate theoretical study, we make the following assumption in this subsection.
	\begin{Assumption}\label{assumpt:G_f}
		The graph $G =(V,E)$ is undirected, and the information propagation along every edge of $G$ is independently distributed according to an unknown continuous distribution with probability density function (pdf) $f$ with mean $\mu$. We also assume that $f$ has infinite support.
	\end{Assumption}
	The continuous distributions commonly used in the literature to model diffusions (for example, \cite{Gomez2010, Gom2011}) satisfy \cref{assumpt:G_f}. 
	
	As a notational convention, we use a lower-case letter (e.g., $f$) to denote the pdf of a continuous probability distribution, and the corresponding capital letter (e.g., $F$) for its cumulative distribution function (cdf). Moreover, for any cdf $F$, we write $\bar{F}=1-F$.
	
	Let $u,v$ be two distinct vertices of a graph $G$. We use $X_{u,v}$ to denote the random variable associated with the time it takes for a piece of information to propagate from $u$ to $v$. Let $f_{u,v}$ be the density function of the associated distribution. Because $G$ is undirected, $f_{u,v}=f_{v,u}$. Consider the case where there are multiple paths from $u$ to $v$, but no edge $(u,v)$ in $G$. The propagation time $X_{u,v}$ is then the minimum of the propagation times along each of these paths. Now we can state the main theorem.
	
	\begin{Theorem} \label{thm:stt}
		Suppose that \cref{assumpt:G_f} holds. For two distinct vertices $u,v$, if there is a path connecting $u,v$ containing a vertex different from both $u$ and $v$, then $f_{u,v}\neq f$. 
	\end{Theorem}
	\begin{IEEEproof}
	See \cref{proof:sec:graph}.
	\end{IEEEproof}

	\Cref{thm:stt} suggests that even if the mean propagation time (as was used in the ITI algorithm) for $X_{u,v}$ is similar to that of a single edge, we can use higher moments to determine if $u$ and $v$ are connected by a single edge (except for certain pathological distributions that share the same moments). In the following, we provide bounds on the moments of $X_{u,v}$ to guide us in our extension of the ITI algorithm to general graphs.
	
	\begin{Lemma} \label{lem:YX}
		Suppose that \cref{assumpt:G_f} holds. Consider the graph $G'$ in which an edge $(u,v)\in E$ is removed, and suppose that $G'$ is connected. Let $Y_{u,v}$ be the propagation time from $u$ to $v$ in $G'$ and its pdf be $h$ (with cdf $H$). Then for all $k\geq 1$, and any $\epsilon_0 > \epsilon_1 > 0$, 
		\begin{align*}
		\E[Y_{u,v}^k] - \E[X_{u,v}^k] \geq \epsilon_1 F((\epsilon_0-\epsilon_1)^{1/k})\bar{H}(\epsilon_0^{1/k}).
		\end{align*}
	\end{Lemma}
		\begin{IEEEproof}
	See \cref{proof:sec:graph}.
	\end{IEEEproof}
	
	In particular, by choosing $\epsilon_0-\epsilon_1$ to be sufficiently large and noting that $\bar{H}(\cdot) > 0$ (notice that $h$ is obtained by taking sum and minimum of distributions with infinite support, and itself has infinite support), \cref{lem:YX} implies that $\E[Y_{u,v}^k] > \E[X_{u,v}^k]$ for all $k\geq 1$. Moreover, if the graph $G$ is not too dense in the sense that $F(1)\bar{H}(1)$ is suitably large, then the difference between the means of $Y_{u,v}^k$ (without an edge between $u$ and $v$) and $X_{u,v}^k$ (with an edge between $u$ and $v$) can be made suitably large by choosing a sufficiently large $k$. This allows us to determine if $(u,v)\in E$ based on empirical observations of the propagation times raised to the $k$-th power. However, we may not observe information cascades from every vertex $u$ in the network, and typically have to rely on information cascades starting at sources other than $u$ and $v$. We have the following bound relating the propagation times to $u$ and $v$ from a distinct source node.
	
	\begin{Lemma} \label{lem:lgb}
		Suppose that \cref{assumpt:G_f} holds. Let $G_{u,v}$ be the union of all the simple paths connecting $u,v$ (i.e., the convex hull, according to \cref{def:conv}) in $G$; and $w\in G$. Then for any $k \geq 1$,
		\begin{align*}
		|\E[X_{w,u}]-\E[X_{w,v}]|^k\leq \E[X_{u,v}^k].
		\end{align*}
	\end{Lemma}
	\begin{IEEEproof}
	See \cref{proof:sec:graph}.
	\end{IEEEproof}
	
	\Cref{lem:lgb} shows that if $\E[X^k_{u,v}]$ is small, then so is the more easily computable $|\E[X_{u,w}]-\E[X_{v,w}]|^k$ (recall that we estimate $\E[X_{u,w}]$ using $T_w(u)$ and the transfer step in the ITI algorithm in \cref{sec:alg}). The reverse implication is not necessarily true, but for algorithmic convenience, the lemma suggests that we use an empirical estimate of the latter term as a proxy for $\E[X^k_{u,v}]$.
	
\subsection{Discussions and implications}
\label{subsec:heu}

We now discuss some implications of the results obtained in \cref{subsec:to}. Consider any pair of distinct nodes $u,v$, and the following cases:
	\begin{enumerate} [(i)]
		\item Suppose that the number of paths between $u$ and $v$ is small (i.e., the graph $G$ is sparse).
		\begin{enumerate}
			\item If $u$ and $v$ are connected by an edge, then the sample mean and moments of the propagation time between $u$ and $v$ approximate well the mean and corresponding moments of a diffusion across a single edge.
			\item On the other hand, if $u$ and $v$ are not connected by a direct edge, then the sample mean and variance of the propagation time between $u$ and $v$ are close to \emph{integer multiples} of the mean and variance of a diffusion across a single edge.
		\end{enumerate} 
		
		\item Suppose that the number of paths between $u$ and $v$ is large (i.e., the graph $G$ is dense). According to \cref{lem:YX} and the discussions thereafter, the existence of an edge between $u$ and $v$ can make both the sample mean and higher moments small relative to the corresponding moment values if such an edge is missing. Therefore, in this case, we may choose to infer that the edge $(u,v)$ exists based on the size of the sample mean and higher moments. However, we should mention that if there are too many paths between $u$ and $v$, then in practice, any distribution-based estimation is prone to errors as demonstrated in the discussion and example below. 
	\end{enumerate} 
	
	Our next example demonstrates that it is almost impossible to determine if there exists an edge $(u,v)$ in the graph $G$ if it is very dense. To show this, we need the following result.
	
	\begin{Lemma} \label{lem:sxa}
		Suppose that $X$ and $Y$ are two continuous random variables on $(0,\infty)$, with cdf $P$ and $H$ respectively. Let $Z = \min\{X,Y\}$. Then the total variation distance between the distributions of $Y$ and $Z$ is bounded from above by $\inf_{\epsilon>0} \{\bar{H}(\epsilon)+P(\epsilon)\}$.  
	\end{Lemma}
	
	\begin{Example}\label{ex:neg}
		Suppose that the propagation along each edge are i.i.d.\ with exponential distribution having mean $1$. Then, we have $F(x) = 1-e^{-x}$ for $x\geq 0$. Assume that there are $k$ independent paths of length $l$ between two distinct nodes $u$ and $v$, which are not connected by an edge. Along each path, the propagation follows a Gamma distribution $\Gamma(l,1)$, whose cdf is $1-e^{-x}\sum_{i=0}^{l-1}x^i/i!$. Let $Y$ be the propagation time between $u$ and $v$. Then, it can be shown that its complementary cdf $\bar{H}(\epsilon) \leq (e^{-\epsilon}\sum_{i=0}^{l-1}\epsilon^i/i!)^k.$ 
		
		Let $Z$ be the propagation time from $u$ to $v$ if the edge $(u,v)$ is added to the graph. By \cref{lem:sxa}, the total variation distance between $Y$ and $Z$ is bounded from above by 
		\begin{align}\label{ex_upper}
		\inf_{\epsilon>0}\left(e^{-\epsilon}\sum_{i=0}^{l-1}\epsilon^i/i!\right)^k+1-e^{-\epsilon}.
		\end{align}
		We have $\lim_{\epsilon \to 0^+}(1-e^{\epsilon}) = 0.$ On the other hand, the derivative of $e^{-x}\sum_{i=0}^{l-1}x^i/i!$ is $-e^{-x}x^{l-1}/(l-1)!<0$ for $x>0.$ This means that $e^{-\epsilon}\sum_{i=0}^{l-1}\epsilon^i/i!<1$ for any $\epsilon>0$. Therefore, if $l$ is fixed, we can always choose $\epsilon$ small enough and $k$ large enough such that the upper bound \eqref{ex_upper} is as close to $0$ as we wish. This suggests that in practice, if there are many paths between the two nodes, then it is almost impossible to determine if there is an edge between them or not by any distribution-based method.
		
		As a specific numerical example, if $l=2$ and $k=100$, \cref{ex_upper} drops below $0.3$.
	\end{Example}
	
	\subsection{The graph inference algorithm} \label{sec:gia}
	
	The discussion in \cref{subsec:heu} can be summarized in the following dichotomy: when the graph is sparse (as measured by the edge to vertex ratio, for example) or the distributions of the propagation times along each edge have small variance, it is enough to use the mean of the distributions as in the case of trees. On the other hand, if the graph is highly connected, the existence of an edge between two vertices $u$ and $v$ can make the mean propagation time between $u$ and $v$ small relative to $\mu\tc{1}$. Therefore, it is instructive to use the length of the propagation time between $u$ and $v$ to decide if they are connected by an edge or not. The same consideration applies to other moments (as compared against unbiased sample moments); and they can be used as additional criteria to decide the existence of edges.   
	
	In a general graph, it might not be known whether the connection between two vertices $u$ and $v$ is dense or sparse. One way to overcome such a difficulty is to compute the difference between the sample mean (respectively, sample moments) with both the theoretical mean (respectively, theoretical moments) as well as $0$. Once these two values are obtained, it is enough to take the smaller one. Hence the weight matrix being used in ITI should be modified based on available moment information. Suppose that the average mean $\mu\tc{1}$ and average second order moment $\mu\tc{2}$ of the propagation time along each edge are known. We define the following:
	\begin{dgroup*}
		\begin{dmath}[label={W_1}]
			W_1(u,v) = \ofrac{|V_c|}\min\left\{ \sum_{v_i\in V_c} |T_{v_i}(u)-T_{v_i}(v)|,  
			\sum_{v_i\in V_c} \left||T_{v_i}(u)-T_{v_i}(v)|-\mu\tc{1}\right|\right\},
		\end{dmath}
		\begin{dmath}[label={W_2}]
			W_2(u,v) = \ofrac{|V_c|}\min\left\{ \sum_{v_i\in V_c} |T_{v_i}(u)-T_{v_i}(v)|^2,  
			\sum_{v_i\in V_c} \left||T_{v_i}(u)-T_{v_i}(v)|^2-\mu\tc{2}\right|\right\},
		\end{dmath}
	\end{dgroup*}
	and
	\begin{align} \label{eq:edge-mod}	
	W(u,v) = W_1(u,v) + W_2(u,v).
	\end{align}
	We then choose $n\cdot \text{deg}_{\text{ave}}/2$ edges with the smallest $W(u,v)$ values to form the estimated graph, where $\text{deg}_{\text{ave}}$ is an estimate of the average degree. In the case of general graphs, we assume that we have some prior knowledge of the underlying network, so that a reasonable estimation of the average degree can be performed (similar to \cite{Gomez2010}). There are quite a few important occasions that we can do so, and we list a few of them as follows:
		\begin{enumerate}[(i)]
			\item We know how the graph is generated, or the distribution that governs the generation of the graph. For example, the graph generated according to the Erd\H{o}s-R\'{e}nyi graph \cite{ER_random} has an expected degree for each vertex. If the network is modeled using the Erd\H{o}s-R\'{e}nyi graph, we can regard the expected degree as the average degree of the graph. Another example is the Forest-fire model \cite{Les07}. If the forward and backward burning	probabilities are available, we can obtain an estimate of the average degree.
			\item There are a small percent of vertices whose degrees are available. For example, in a social network, we can take a survey to learn the number of friends of some users and estimate the average degree $\text{deg}_{\text{ave}}$ through sampling.
			\item \label[occasion]{case:3}
			There are situations where \emph{extreme value theory} \cite{Castillo1988} can be applied and a direct estimation of the average degree from the cascades is possible. As a typical example, suppose that the propagation times along all edges are i.i.d.\ with exponential distribution having mean $\mu\tc{1}$. Consider a cascade with source vertex $u$ whose degree is  $\text{deg}(u)$, define
			\begin{align*}
			t_{\text{min}}(u)=\min_{v\in V\backslash u}T_u(v).
			\end{align*}
			It is easy to verify that $t_{\text{min}}(u)$ follows the exponential distribution with mean $\mu\tc{1}/\text{deg}(u)$. Therefore, we can estimate the degree of $u$ as $\mu\tc{1}/t_{\text{min}}(u)$. Since we observe a collection of cascades, then by averaging we obtain the estimate 
			\begin{align}\label{eq:deg_avg}
			\text{deg}_{\text{ave}}=\frac{\sum_{u\in V_c}\mu\tc{1}/t_{\text{min}}(u)}{|V_c|}.
			\end{align}			
		Because of the outlier problem when estimating the parameter of an exponential distribution, we revise our estimation to make it robust according to \cite{Gather1986,Ahmed2005}. For other spreading models, we adopt the same heuristic to obtain $\text{deg}_{\text{ave}}$. Simulation results in \cref{subsec:general_graphs} demonstrates that \cref{eq:deg_avg} is a reasonable estimate of the average degree. 
		\end{enumerate}

Summarizing the above discussions, our Graph Inference (GI) algorithm as a heuristic extension of ITI is shown in \cref{algo:gi}.	
	
	\begin{algorithm}[!htb]
		\caption{Graph Inference (GI) algorithm} \label{algo:gi}
		\begin{algorithmic}[1]		
			\STATE {\bf Input} $I:$ number of transfer steps, $m_s:$ number of edges selected for the selection step, $\text{deg}_{\text{ave}}:$ estimated average degree of $G$, $V_c$: set of cascade sources, $\{T_{v_i}(\cdot): v_i\in V_c\}$: timestamp information, the average mean $\mu\tc{1}$ and second order moment $\mu\tc{2}$ of the propagation time.%
			\STATE Initialize $\eta_{v}=0$ for each $v \in V_c$.
			\FOR {$s=1,\ldots,I$}
			\STATE Compute weight matrix $W$ in \eqref{eq:edge-mod} and select $m_s$ pairs $\left\lbrace (v_i,v_j)\right\rbrace $ having the $m_s$ smallest $W(v_i,v_j)$ weights.\\
			\FOR {each pair of $(v_i,v_j)$ selected}
			\STATE Construct $T_{v_i}^{v_j}(\cdot)$ and $T_{v_j}^{v_i}(\cdot)$ using \eqref{cas-gen}. 
			\STATE Update $\eta_v=\eta_v+1$ for $v=v_i,v_j.$
			\STATE Update $T_{v_i}(\cdot) = (\eta_{v_i}T_{v_i}(\cdot) + T_{v_i}^{v_j}(\cdot))/(\eta_{v_i}+1)$ and $T_{v_j}(\cdot) = (\eta_{v_j}T_{v_j}(\cdot) + T_{v_j}^{v_i}(\cdot))/(\eta_{v_j}+1)$.
			\ENDFOR
			\ENDFOR 
			\STATE Select $n\cdot \text{deg}_{\text{ave}}/2$ edges $\left\lbrace (u,v)\right\rbrace$ corresponding to the smallest $W(u,v)$ weights.
			\STATE {\bf Output} $\left\lbrace (u,v)\right\rbrace$.
		\end{algorithmic}
	\end{algorithm}  
	
	A possible generalization of \cref{algo:gi} if higher-order moments are available is to modify $W_1, W_2$ and hence $W$ accordingly as follows. According to \cref{lem:lgb} and the discussion thereafter, we may use $|T_{v_i}(u)-T_{v_i}(v)|$ to estimate the sample moments. For each $k\geq 1$, the $k$-th sample moment is denoted by $S^{(k)}_{u,v}(v_i)$ as the average of $|T_{v_i}(u)-T_{v_i}(v)|^k$ over $v_i\in V_c$. Suppose the $k_1=1,k_2,\ldots,k_m$-th moments are available. Let $\phi$ be a continuous $m$-variable function, and for each $1\leq i\leq m$, let
	\begin{dmath}
		W_i(u,v) = \ofrac{|V_c|}\min\left\{ \sum_{v_i\in V_c} S^{(k_i)}_{u,v}(v_i), 
		\sum_{v_i\in V_c} \left|S^{(k_i)}_{u,v}(v_i)-\mu\tc{k_i}\right|\right\}.
	\end{dmath}
	We then define 
	\begin{align}\label{eq:edge-recg}
	W(u,v) = \phi(W_1(u,v),\ldots,W_m(u,v)).
	\end{align}
	Under this generalization, the procedure depicted in \cref{algo:gi} uses $m=2$, $k_1=1$, and $k_2=2$, while $\phi$ is the averaging function. The choice of $\phi$ should reflect one's belief about which moment should play a more important role in the network inference task.
	
	\section{Simulation results} \label{sec:sim}
	
	In this section, we present simulation results to illustrate the performance of our proposed topology inference algorithms. We first apply our ITI algorithm on tree networks, and compare its performance with the tree reconstruction (TR) algorithm proposed in \cite{Abr2013}, which is most similar to ours in assumptions. We then perform simulations on general graphs, including some real-world networks, and compare the performance of the GI algorithm with the NetRate algorithm proposed in \cite{Gom2011}. As the NetRate algorithm assumes knowledge of the diffusion distribution, we also study the performance impact of a mismatch between the assumed and actual distributions. To the best of our knowledge, there are no other works on topology inference making similar assumptions as ours. TR and NetRate are the closest methods that allow feasible comparison. 
	
	Suppose that $A_G$ is the true adjacency matrix of $G$ and $A$ is an estimated adjacency matrix. To evaluate the performance of our method, we define the \emph{edge recovery rate} as 
	\begin{equation}
	\label{eq:rate}
	R\triangleq1-\frac{\sum_{1\leq i<j\leq n}|A(i,j)-A_G(i,j)|}{2|E|}.
	\end{equation}
	For the same number of edges, each mistake in identifying an edge causes a mistake at another pair of vertices. To account for this, we have a factor of $2$ in the denominator; and this makes $R$ a real number in $[0,1]$. The term $1-R$ is called the \emph{error rate}, in which both undetected edges and false positives are taken into account. 
	All the simulation results shown in the following sections are averaged over 200 trials. For each trial, given the graph $G$, we randomly pick $|V_c|$ vertices. For each source vertex $v_i\in V_c$, we initiate the diffusion process and obtain a cascade. We then run the algorithms given the $|V_c|$ cascades to obtain the edge recovery rate for this trial.
	
	\subsection{Tree networks}\label{Subsec:sim_tree}
	
%

	\begin{figure}[!htb]
		\subfigure[]{
			\begin{minipage}[b]{0.5\linewidth}
				\centering
				\includegraphics[height=6cm]{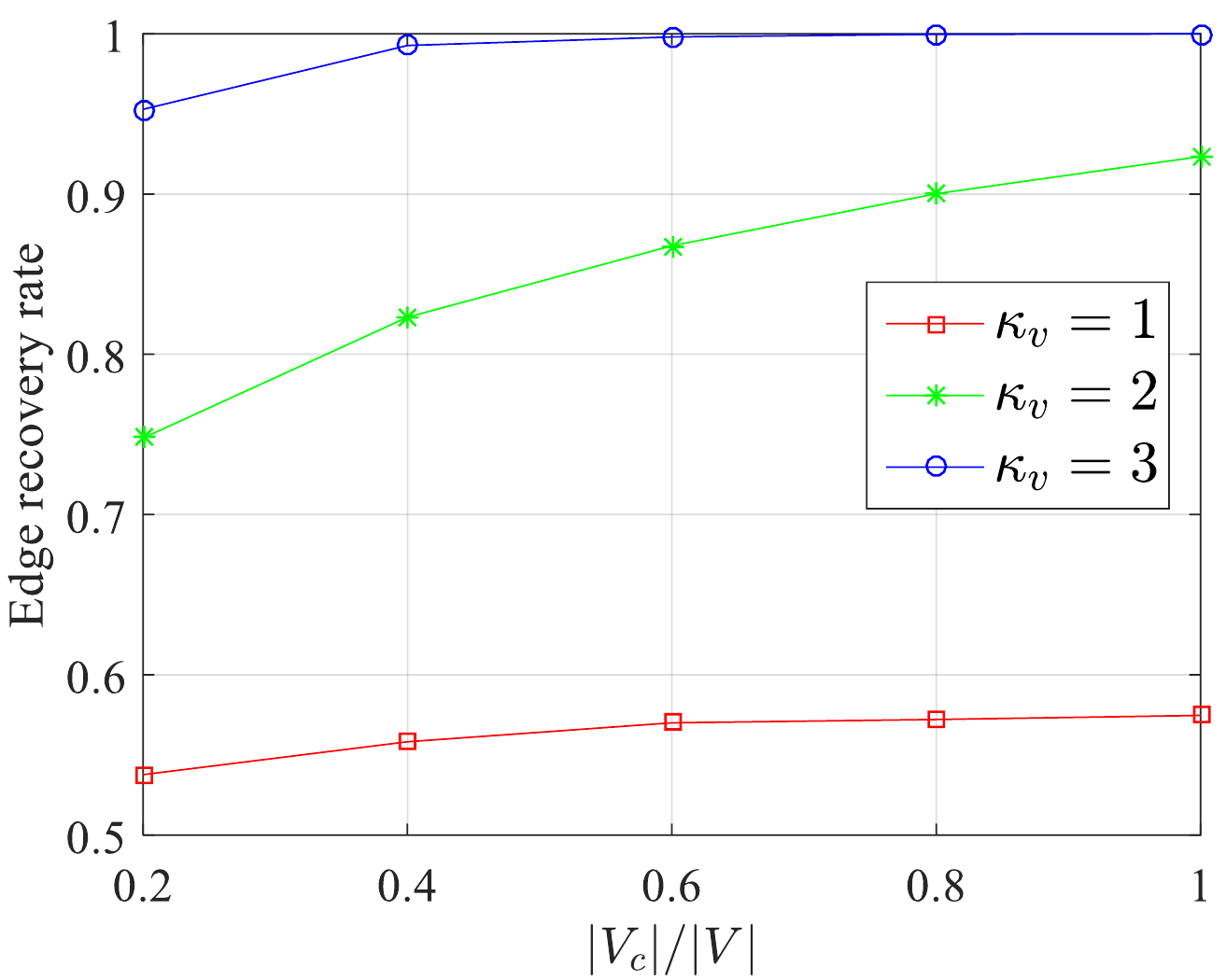}
			\end{minipage}
			\label{fig:com}
		}%
		\subfigure[]{
			\begin{minipage}[b]{0.5\linewidth}
				\centering
				\includegraphics[height=6cm]{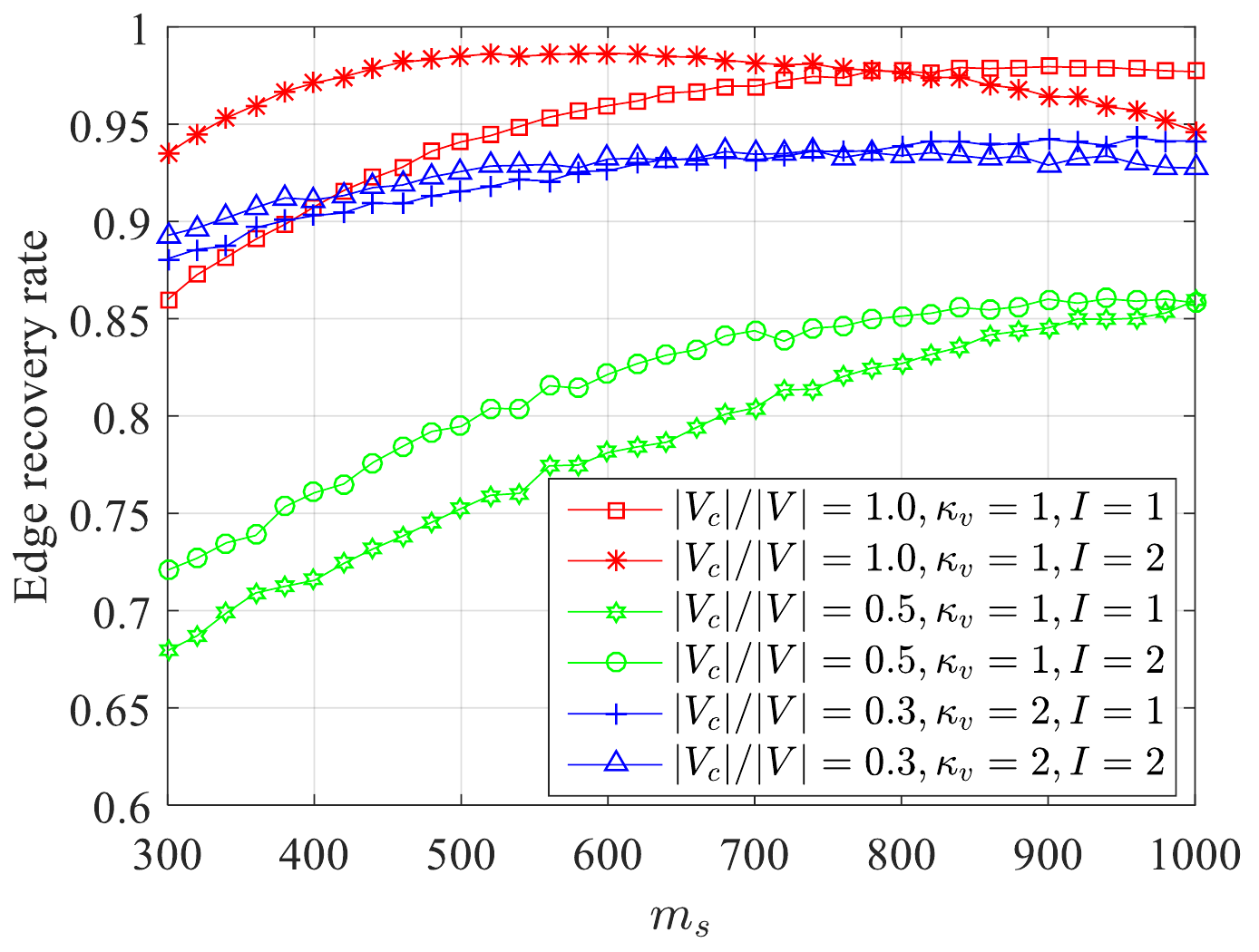}
			\end{minipage}
			\label{fig:six}
		}
	\caption{\blue{This figure shows the performance of ITI for different parameters.} In (a), performance of ITI with $I=0$ with varying $|V_c|$ and $\kappa_v$. In (b), performance of ITI with varying $|V_c|, \kappa_v, I$ and $m_s$. Curves with the same color have the same $|V_c|$ and $\kappa_v$ values.}
	\label{figure:iti}
    \end{figure}

	In this subsection, we show and discuss simulation results to study the performance of our proposed algorithm for tree networks. As we are considering tree networks, in the reconstruction step, we always have $|E|=|V|-1$ edges. 
	
	In each simulation run, we randomly generate trees with $|V|=500$ vertices. We then randomly choose $|V_c|$ distinct sources, and generate $\kappa_v$ cascades per source using independent exponential spreading with mean $\mu\tc{1}=1$ at each edge. In applications, $\kappa_v$ can be different for distinct nodes $v$; however, in our simulations below, we keep it constant for all $v$ for simplicity. We apply the ITI algorithm, and evaluate the performance by using \eqref{eq:rate}. 
	
	We first study the edge recovery rate by varying $|V_c|$ and $\kappa_v$, and skip the selection and transfer steps by setting $I=0$ (see \cref{fig:com}). We notice that $\kappa_v$ has a significant impact on the performance. If the number of cascades per source $\kappa_v$ is small, the performance can be greatly improved if we perform additional transfer steps (see \cref{fig:six}). These results demonstrate the usefulness of the theory developed in \cref{sec:re}. The choice $I=2$ is usually enough, and the number of edges $m_s$ in the selection step can be chosen between $|V|$ and $1.5|V|$. Moreover, comparing the curves for $|V_c|/|V|=0.3, \kappa_v=2$ with $|V_c|/|V|=0.5, \kappa_v=1$, we see that the performance is improved if we have more cascades originating from the same source. This suggests that the averaging process allows a better approximation of the distance on a tree. 
	
\begin{figure}[!htb]
		\centering 
		\begin{minipage}[b]{0.5\linewidth}
			\centering
			\includegraphics[height=5.6cm]{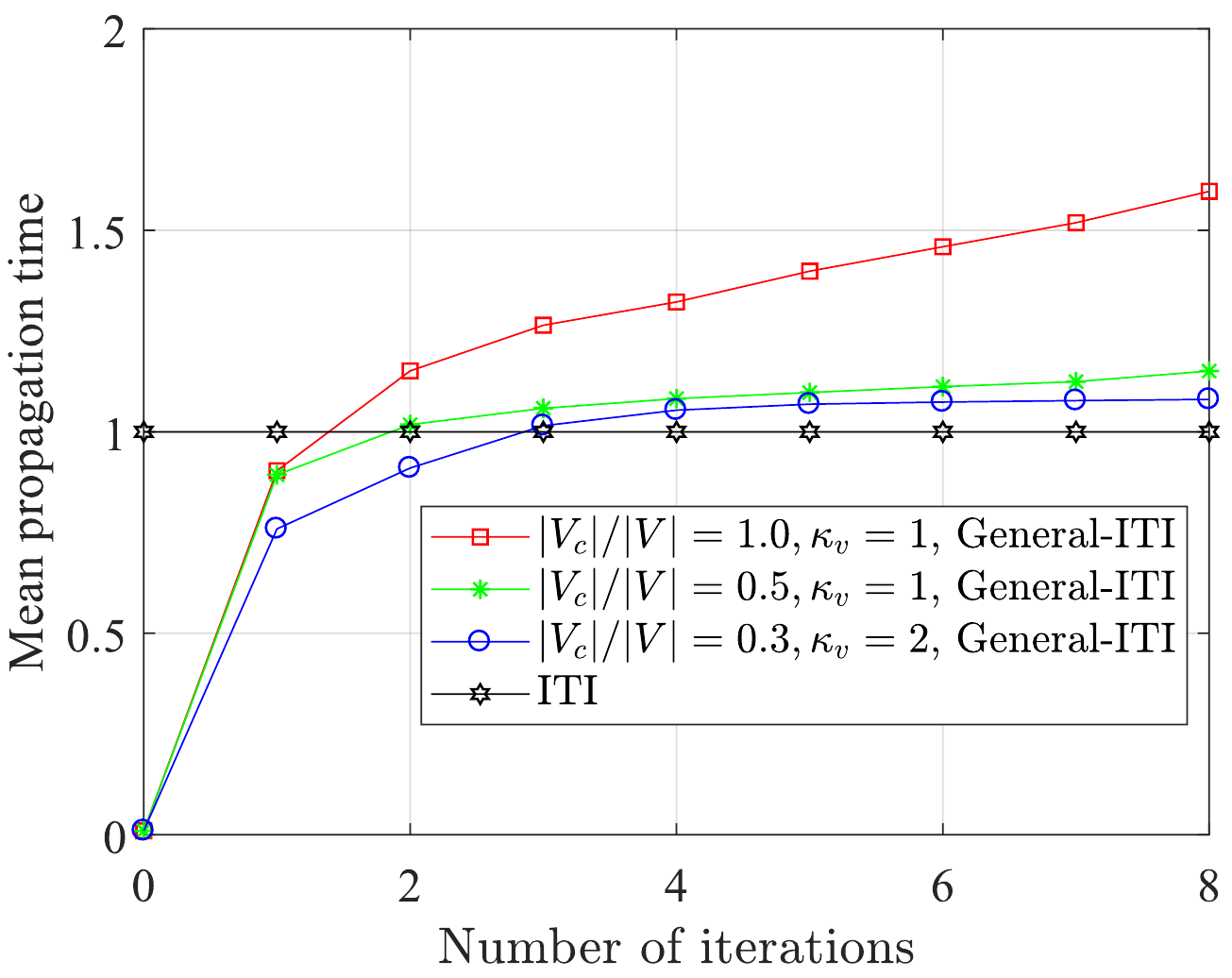}
			\centerline{(a)}
		\end{minipage}%
		\begin{minipage}[b]{0.5\linewidth}
			\centering
			\includegraphics[height=5.6cm]{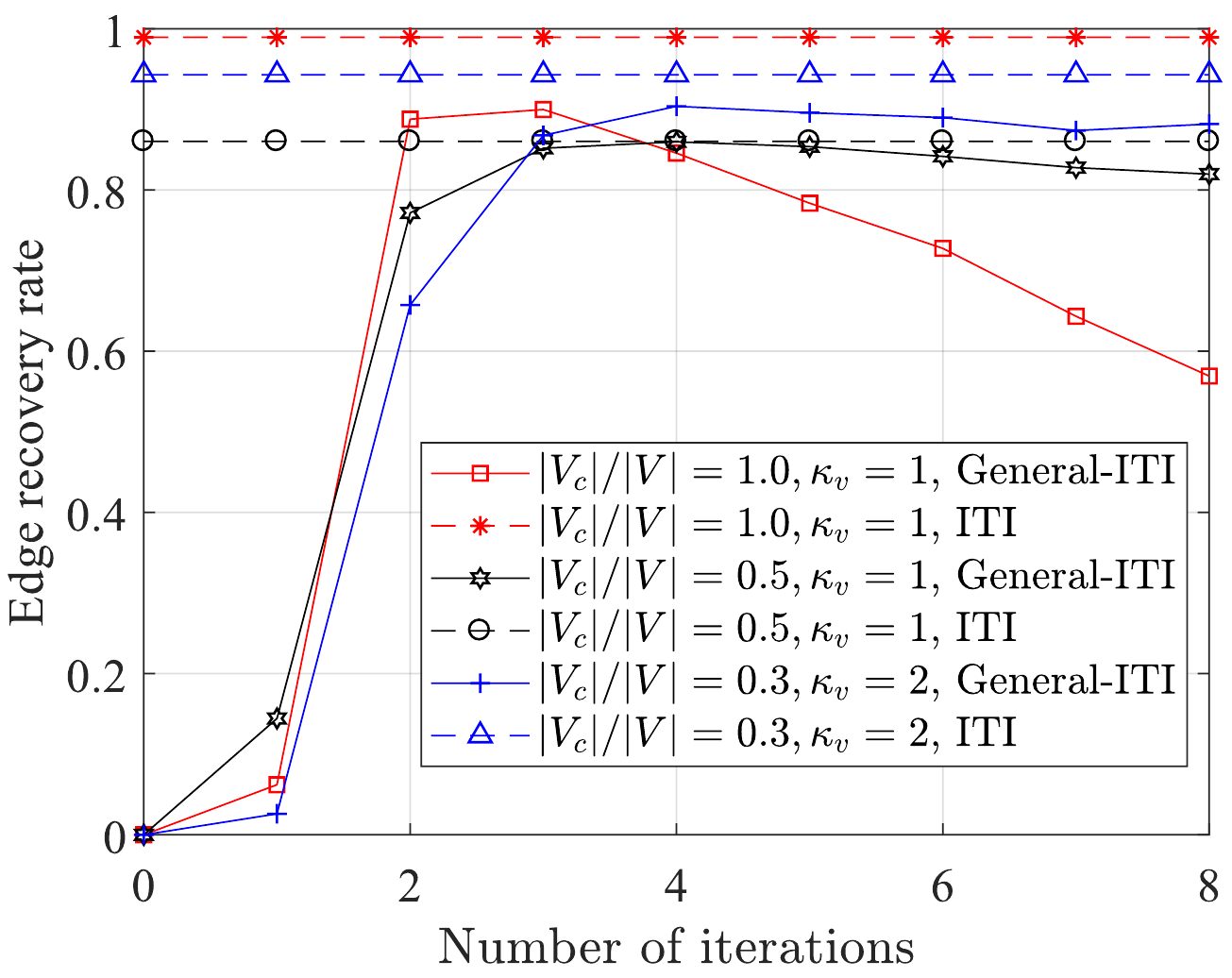}
			\centerline{(b)}
		\end{minipage}
		\caption{Performance comparison between General-ITI and ITI.} \label{fig:GITI}
	\end{figure}

In our experiments, we found that it is still possible to infer a tree topology when the mean propagation time is unknown using the following procedure: We first initiate $\mu\tc{1}$ to be a small value and run ITI to obtain a collection of edges, each with an estimated propagation time. We then estimate the mean propagation time by averaging the propagation times along the estimated edges. This procedure is then repeated. We call this heuristic method General-ITI. We compare the performance of General-ITI with ITI ($I=2$ and $m_s=1.5|V|$) in \cref{fig:GITI}. We see that the estimated mean propagation time increases and exceeds the actual value as the number of iterations increases. The recovery performance improves over 2 or 3 iterations and is comparable with ITI, where the actual mean is known.
	
\begin{figure}[!htb]
		\centering 
		\begin{minipage}[b]{0.5\linewidth}
			\centering
			\includegraphics[height=5.6cm]{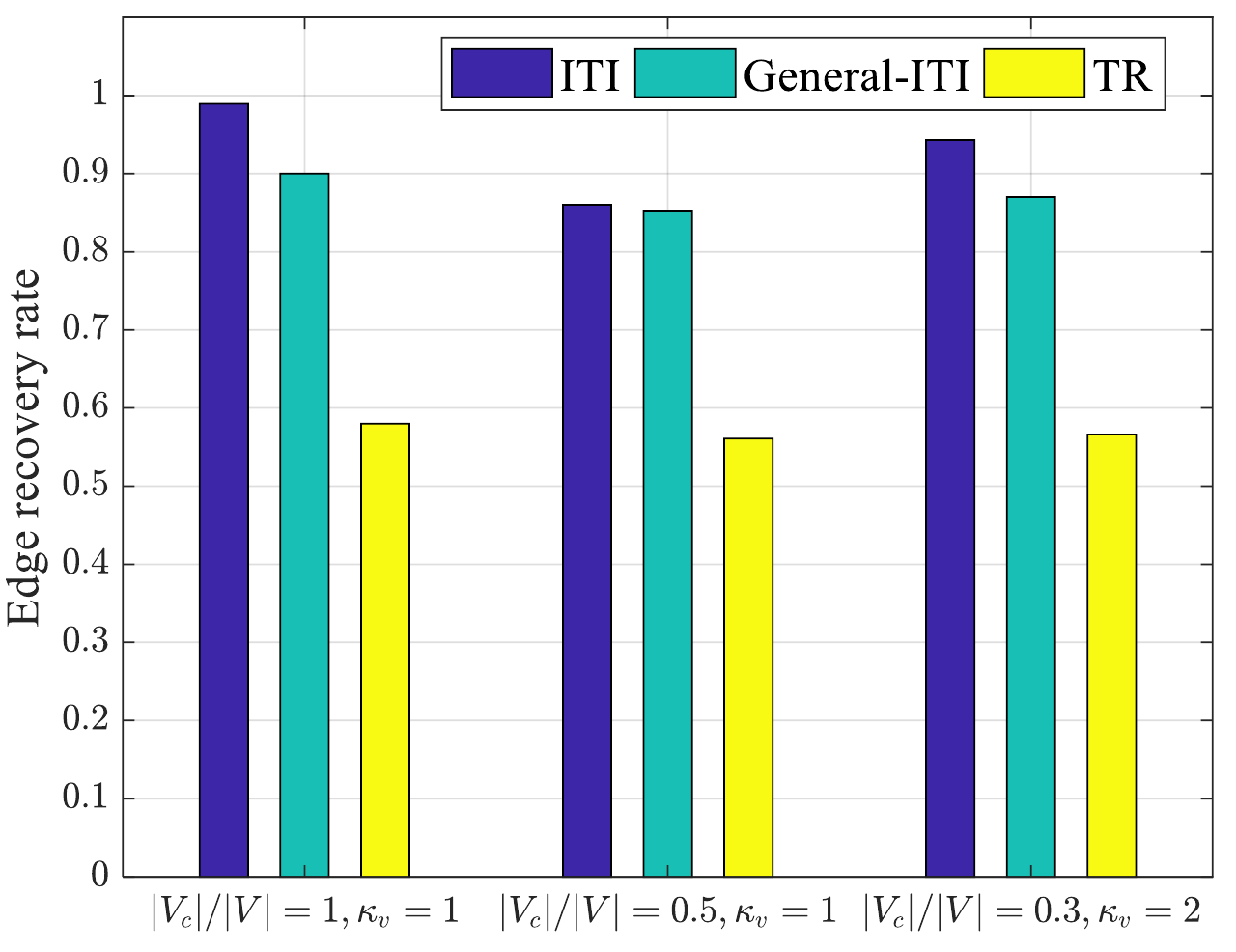}
			\centerline{(a)}
		\end{minipage}%
		\begin{minipage}[b]{0.5\linewidth}
			\centering
			\includegraphics[height=5.6cm]{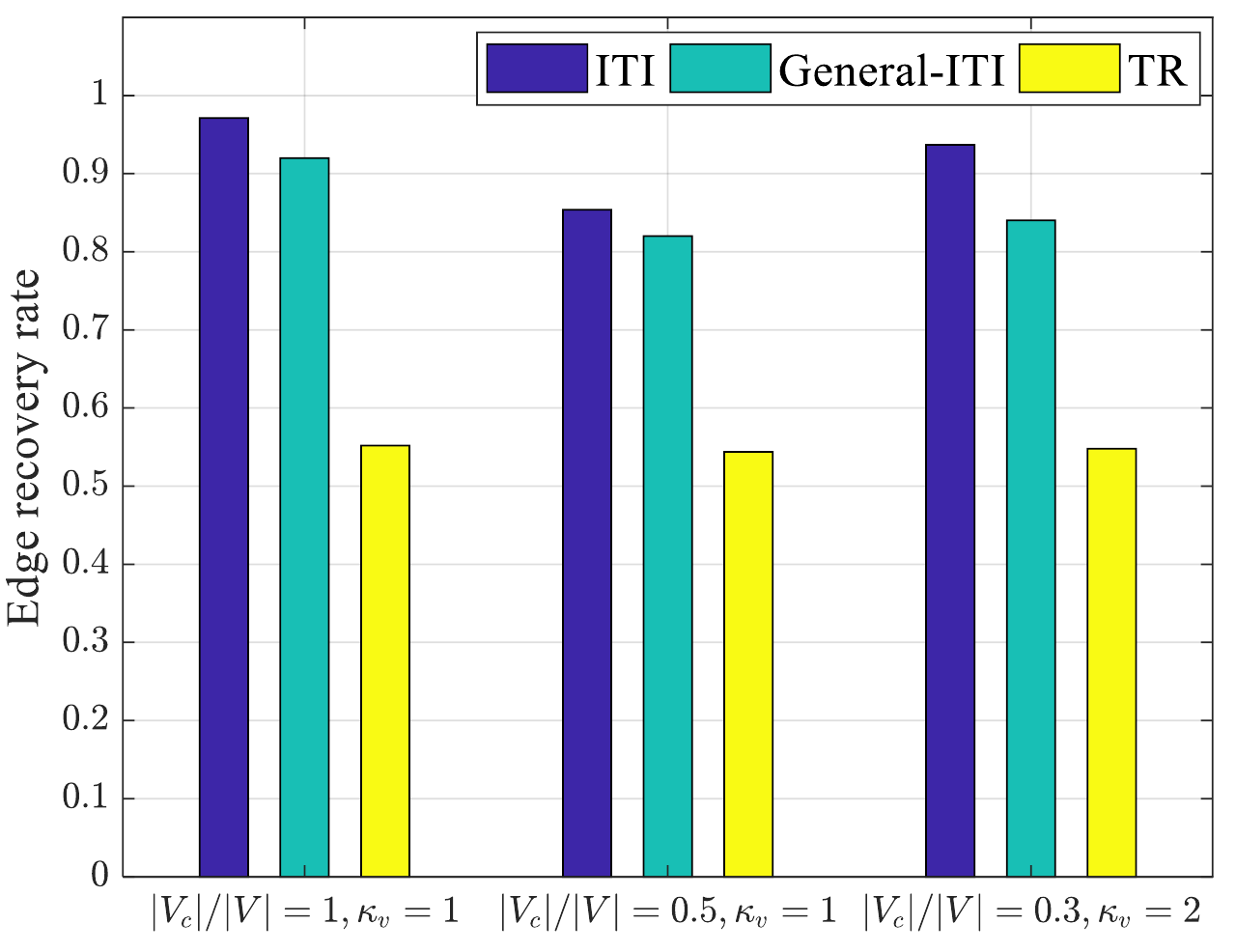}
			\centerline{(b)}
		\end{minipage}
		\caption{Performance comparison between ITI, General-ITI and TR. In (a),  the propagation time follows a fixed exponential distribution. In (b), the propagation time along each edge follows a randomly selected Gamma distribution.} \label{fig:ITI_TR}
	\end{figure}

	Finally, we compare ITI and General-ITI (3 iterations) with the TR algorithm proposed in \cite{Abr2013}. Although the theoretical part of \cite{Abr2013} assumes i.i.d.\ exponential distribution for propagation time along different edges, the TR algorithm itself requires $|V|$ and $\{T_{v_i}(\cdot): v_i \in V_c\}$ as the only inputs. The comparison is shown in \cref{fig:ITI_TR}. In \cref{fig:ITI_TR}(b), we test the effect of having different diffusion distributions for different edges. The distributions are randomly selected unknown Gamma distributions with the same mean. As shown from the plots, our methods perform much better in all the cases because our methods do not require identical distributions for propagation times along different edges. The performance of General-ITI suggests that \blue{we can even recover a tree network} without knowing the exact mean propagation time.

	\subsection{General graphs}\label{subsec:general_graphs}

	For general graphs, we first perform experiments to show the estimation accuracy of $\text{deg}_{\text{avg}}$ if we adopt \cref{eq:deg_avg} proposed in \cref{sec:gia} \cref{case:3}. We consider two graphs: the Forest-fire network \cite{Les07}, and a real-world Email network \cite{Les07} and four spreading models for the propagation time along each edge: exponential distribution Exp$(1)$ with fixed mean 1, exponential distribution with mean chosen uniformly and randomly in $[0.5,1.5]$ (denoted as Exp$([0.5,1.5])$), Gaussian distribution $\calN(1,0.5^2)$ with mean 1 and deviation 0.5, and Gamma distribution $\Gamma(1,2)$ with shape parameter 1 and scale parameter 2. For each simulation, we choose $|V_c|/|V|$ uniformly at random from $[0.1, 1.0]$. We compute the sample mean $\mu_r$ and sample standard deviation $\sigma_{r}$ of $\text{deg}_{\text{ave}}/\text{deg}^*_{\text{ave}}$ using 200 trials where $\text{deg}^*_{\text{ave}}$ is the actual average degree of the graph. Simulation results are shown in \cref{tab:deg_avg}. We see that \cref{eq:deg_avg} gives the best estimate in the exponential spreading model. For other three models, we are also able to estimate the average degree within a reasonable range. In all our subsequent simulations, we use \cref{eq:deg_avg} to obtain $\text{deg}_{\text{ave}}$.
	\begin{table}[!htb]
		\caption{Performance of the estimated average degree.}
		\centering  
		\begin{tabular}{|l|c|c|c|c|}  
			\hline
			$\mu_r,\ \sigma_{r}$ & Exp$(1)$ & Exp$([0.5,1.5])$ & $\calN(1,0.5^2)$ & $\Gamma(1,2)$\\ 
			\hline  
			Forest-fire network &0.98, 0.2 &1.09, 0.22 &1.24, 0.28 &0.98, 0.18  \\
			\hline        
			Email network&1.14, 0.17 &1.24, 0.23 &0.91, 0.37 &1.20, 0.21 \\   
			\hline
		\end{tabular}
		\label{tab:deg_avg}
	\end{table}
	
We compare our GI algorithm with the NetRate algorithm proposed in \cite{Gom2011}\footnote{The source code for NetRate was retrieved from SNAP (Stanford Network Analysis Project; \url{http://snap.stanford.edu/data/memetracker9.html}). We thank the authors of \cite{Gom2011} for sharing it online.} and KernelCascade algorithm proposed in \cite{Dunips}. NetRate works under a completely different set of assumptions (in particular, \cite{Gom2011} assumes propagation along edges follows one of the following families of distributions: exponential, power-law, or Rayleigh). As it performs better than NetInf in \cite{Gomez2010} and ConNie in \cite{MyeLes2010}, we do not compare against the latter two methods. For KernelCascade, we choose a set of hyperparameters and kernels similar to the settings in \cite{Dunips}.

\begin{figure}[!htb]
\centering 
\begin{minipage}[b]{0.5\linewidth}
	\centering
	\includegraphics[height=5.6cm]{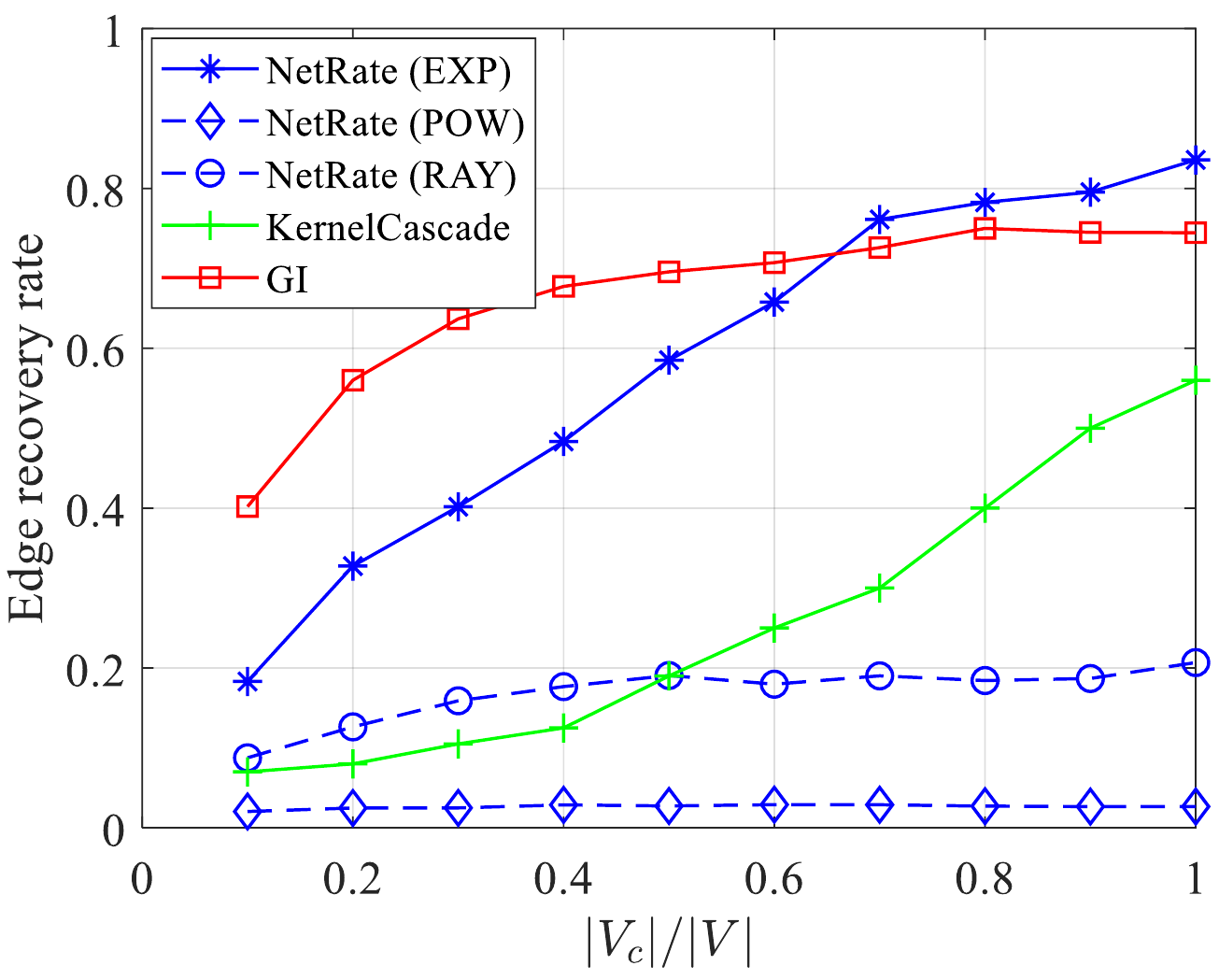}
	\centerline{(a)}
\end{minipage}%
\begin{minipage}[b]{0.5\linewidth}
	\centering
	\includegraphics[height=5.6cm]{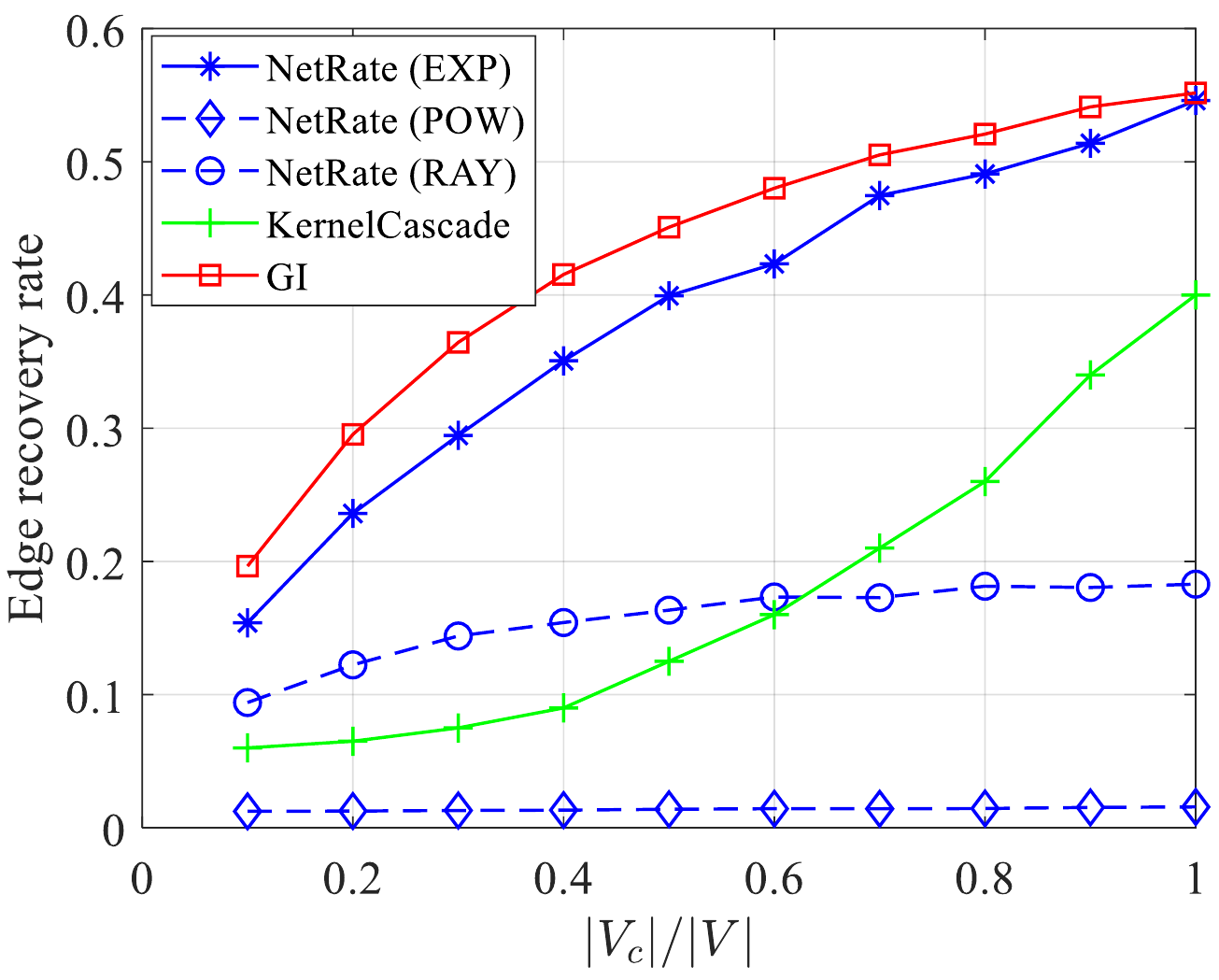}
	\centerline{(b)}
\end{minipage}
\begin{minipage}[b]{0.5\linewidth}
	\centering
	\includegraphics[height=5.6cm]{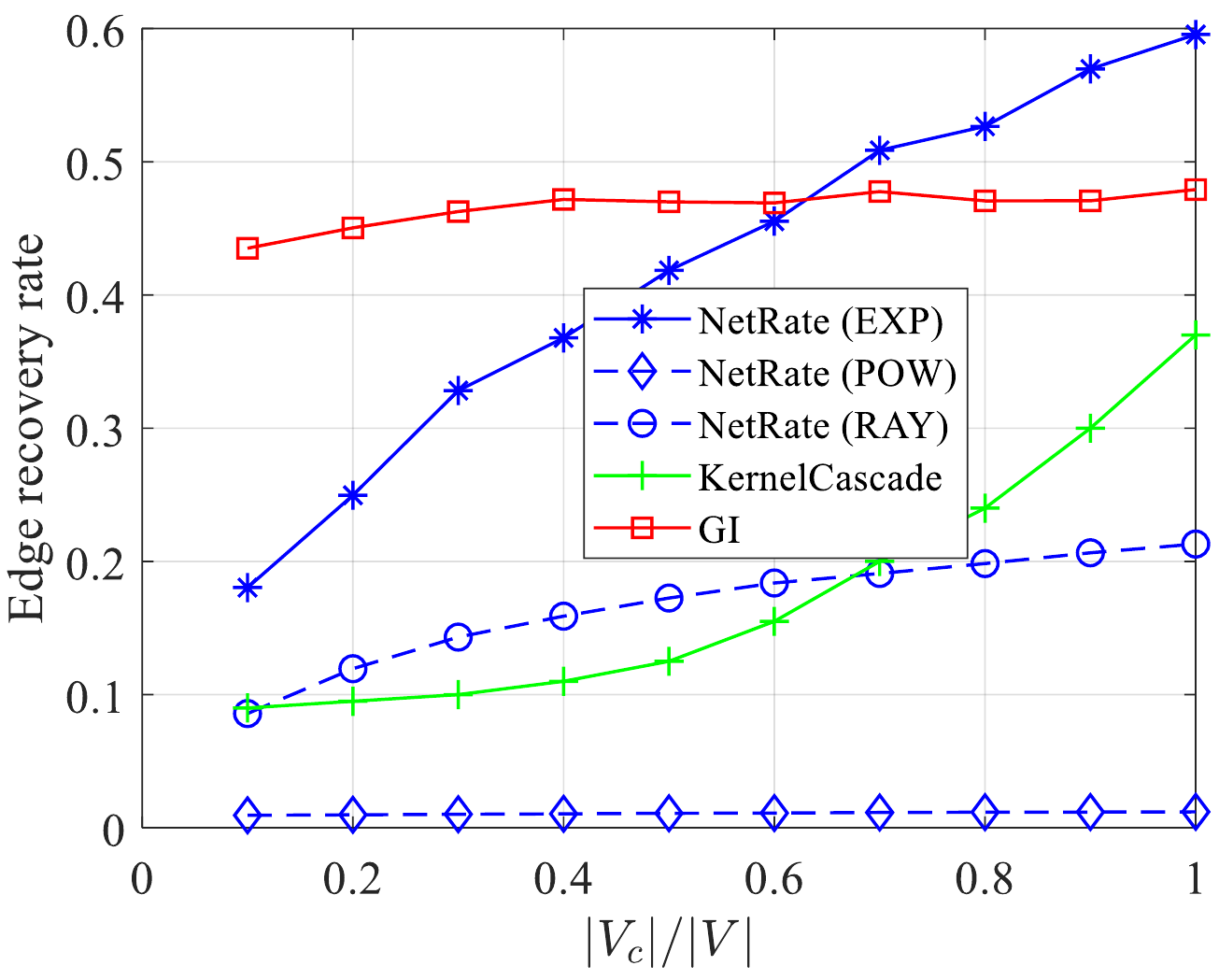}
	\centerline{(c)}
\end{minipage}%
\begin{minipage}[b]{0.5\linewidth}
	\centering
	\includegraphics[height=5.6cm]{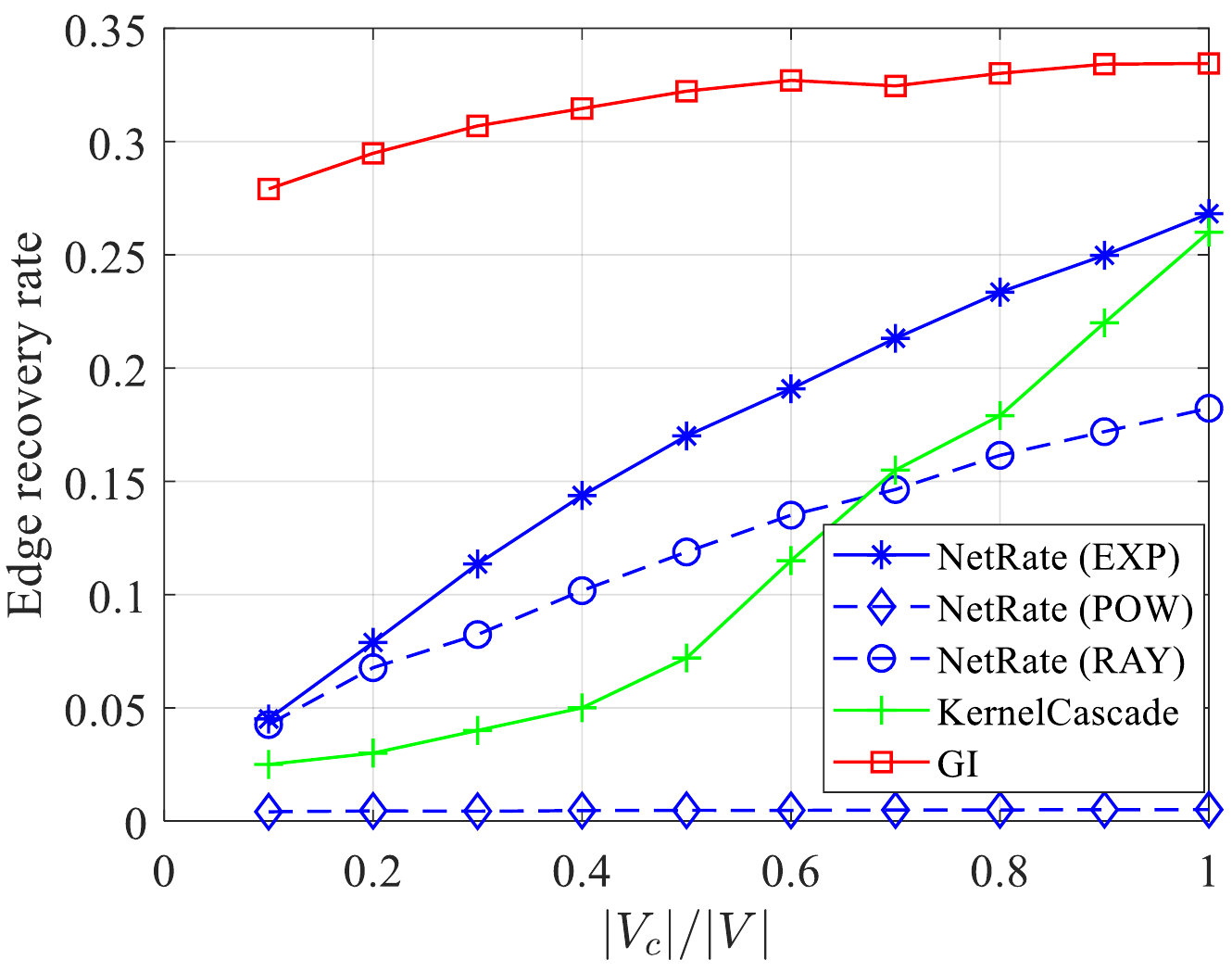}
	\centerline{(d)}
\end{minipage}
\caption{Performance comparison between the NetRate algorithm (blue), KernelCascade algorithm (green) and the GI-algorithm (red), on (a) E-R graphs with $300$ nodes and average degree $4$, (b) E-R graphs with $300$ nodes and average degree $8$, (c) Forest-fire network with $500$ nodes and average degree about $5$ and (d) Email network with $500$ nodes and average degree $12$. The information propagation along the edges follow the exponential distribution with parameter $\lambda=1$. The dashed curves show the performance of NetRate if there is a distribution mismatch for comparison purpose.} \label{fig:methods_4}
\end{figure}
	
To accommodate comparison with NetRate, we use the standard exponential distribution with mean $1$ for propagation time along each edge. We compare the performance of NetRate, KernelCascade and GI on Erd\"{o}s-R\'{e}nyi graphs, the Forest-fire network \cite{Les07}, and a real-world Email network \cite{Les07}. The parameters of all networks are described in the plots.

From \cref{fig:methods_4}, we see that our method performs best in all the tested cases if the number of cascades does not exceed $60\%$ of the number of nodes. For ER-graphs with large average degree and the Email network, GI performs best for the entire spectrum of $|V_c|/|V|$ from $10\%$ to $100\%$. GI has a noticeably better performance than NetRate and KernelCascade when $|V_c|/|V|$ is very small. For example, in the case of the Email network and $|V_c|/|V|=10\%$, NetRate and KernelCascade have less than $5\%$ edge recovery while GI has more than $27\%$ edge recovery. We note that since the Email network is dense, all inference methods based on the assumption that diffusion across each edge follows a distribution will have limited performance (cf.\ \cref{ex:neg}). On the other hand, the advantage of NetRate starts to show up when there is a large number of cascades. In particular, if the ratio $|V_c|/|V|$ is closer to $1$ (i.e., on average each node sends a cascade), then NetRate has a better performance for certain graph types (\cref{fig:methods_4}(a) and (c)). KernelCascade performs worst but its performance improves  significantly when the number of cascades is large. That is because KernelCascade learns the edge transmission functions from the data, which is difficult to accomplish with limited cascades. In addition, the choice of kernels and hyperparameters, which may vary for different networks or number of cascades, can also affect the performance.
	
	Another advantage of our method is that it is much more computationally efficient than NetRate and KernelCascade. Using the same computational resource (Processor: Intel(R) Xeon(R) CPU E3-1226 v3 3.30GHz, RAM: 8.00GHz) under the same simulation settings, the average time used to run an instance of GI, NetRate and KernelCascade is shown in \cref{tab:ctio}. Our method is more suitable in time-critical applications when computational resources are limited.
	
	\begin{table}[!ht]
		\centering
		\caption{Computation time (in \emph{seconds}) of GI, NetRate and KernelCascade for varying $|V_c|/|V|$}
		\centering
		\scalebox{0.7}{
			\begin{tabular}{r | c c c c c c c c c c}
				\hline\hline
				$|V_c|/|V|$ & $0.1$ & $0.2$ & $0.3$ & $0.4$ & $0.5$ & $0.6$ & $0.6$ & $0.8$ & $0.9$ & $1.0$\\ [0.5ex]
				\hline
				Email (GI) & $11.9$ & $16.7$ & $20.6$ & $25.3$ & $30.0$ & $35.3$ & $38.1$ & $41.8$ & $42.6$ & $44.8$\\
				Email (NetRate) & $866.2$ & $1086.4$ & $1380.5$ & $1730.9$ & $2102.4$ & $2470.6$ & $2877.2$ & $3359.8$ & $3880.1$ & $5263.6$\\
				Email (KernelCascade) & $1206.6$ & $1550.1$ & $1938.2$ & $2504.2$ & $3072.4$ & $4034.1$ & $5031.4$ & $6584.6$ & $8297.7$ & $10852.3$\\
				\hline
				Forest-fire (GI) & $10.1$ & $13.4$ & $15.8$ & $18.9$ & $22.1$ & $25.1$ & $27.4$ & $30.2$ & $33.8$ & $36.5$\\
				Forest-fire (NetRate) & $681.5$ & $1013.3$ & $1354.3$ & $1690.8$ & $2083.2$ & $2406.7$ & $2806.8$ & $3257.1$ & $3757.4$ & $5128.1$\\
				Forest-fire (KernelCascade) & $739.7$ & $1009.7$ & $1319.8$ & $1731.4$ & $2288.6$ & $3025.1$ & $3788.6$ & $5385.4$ & $6936.3$ & $9534.5$\\
				\hline
		\end{tabular}}
		\label{tab:ctio}
	\end{table}
	
	\subsection{Distribution mismatch}\label{subsec:mismatch}
	
	For the next set of simulations, we test the effect of distribution mismatch on NetRate. Note that since both KernelCascade and GI does not assume any spreading distribution, they do not have such a problem. 
	
	The setup of the simulations is as follows: on each type of network, we generate the cascades using the exponential distribution with rate $1$. We run the NetRate algorithm with the incorrect distributions: either the power-law distribution (POW) or the Rayleigh distribution (RAY), as these are the other two families of distributions discussed in \cite{Gom2011}. The results are shown in the same \cref{fig:methods_4} by dashed curves. 
	
	From the plots, we see that the performance of NetRate drops significantly if there is a distribution mismatch. For example, if the power-law distribution is used, the edge recovery of NetRate is close to $0\%$ regardless of $|V_c|/|V|$ and the network type.
	
	The experiments suggest that prior knowledge of the diffusion distribution is important to guarantee the performance of NetRate. In contrast, for both GI and KernelCascade, no such prior knowledge is required. 
	
	\subsection{Heterogeneous spreading distributions}\label{subsec:Heterogeneous} 
	
%
	
	We next study the performance of GI when the spreading distributions are heterogeneous. Along each edge, the propagation time follows an exponential distribution with mean chosen uniformly and randomly in $[0.5,1.5]$.  
	
	Since both GI and NetRate outperform each other in different $|V_c|/|V|$ regimes, we develop a procedure to fuse their results together. We first run NetRate to estimate the average mean and second order moment of the edge propagation time. Then, we use these in GI to compute the weights $W(u,v)$ in (8). To interpret these weights as ``likelihood scores'', let $W^*= \min_{u,v} W(u,v)$, and
	\begin{align*}
		\ell(u,v) = \phi\left(\frac{W(u,v)-W^*}{\hat{\sigma}}\right),
	\end{align*}
	where $\phi(\cdot)$ is the pdf of the standard normal distribution, and $\hat{\sigma}^2=\ofrac{n_e}\sum_{u,v} |W(u,v)-W^*|^2$ with the sum being taken over the $n_e=\text{deg}_{\text{ave}}\cdot n/2$ smallest $W(u,v)$.  We treat $\ell(u,v)$ as the likelihood score for $(u,v)$. The intuition is that those $(u,v)$ with $W(u,v)=W^*$ are most likely edges, and we take this as the baseline. The ``likelihood'' of any other $(u,v)$ being an edge is then computed w.r.t.\ this baseline. NetRate can also produce such likelihood scores (or in equivalent forms) for distinct nodes being connected by an edge in the graph $G$. We first normalize the scores (with unit total sum) for both methods; and take their average as the final likelihood scores. The edges with higher scores are selected. We call this approach NetRate-GI.
	
\begin{figure}[!tb]
		\centering 
		\begin{minipage}[b]{0.5\linewidth}
			\centering
			\includegraphics[height=5.8cm]{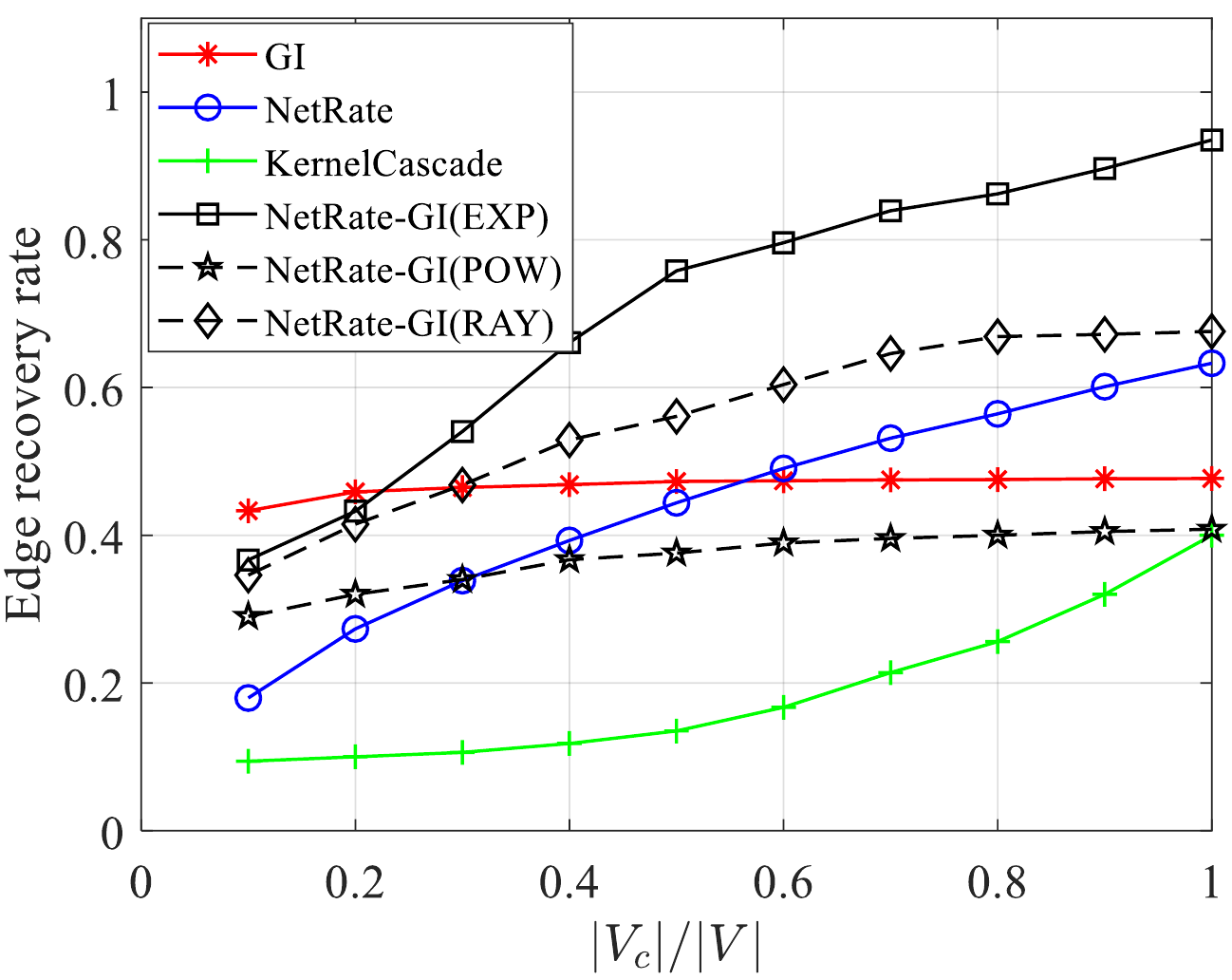}
			\centerline{(a)}
		\end{minipage}%
		\begin{minipage}[b]{0.5\linewidth}
			\centering
			\includegraphics[height=5.8cm]{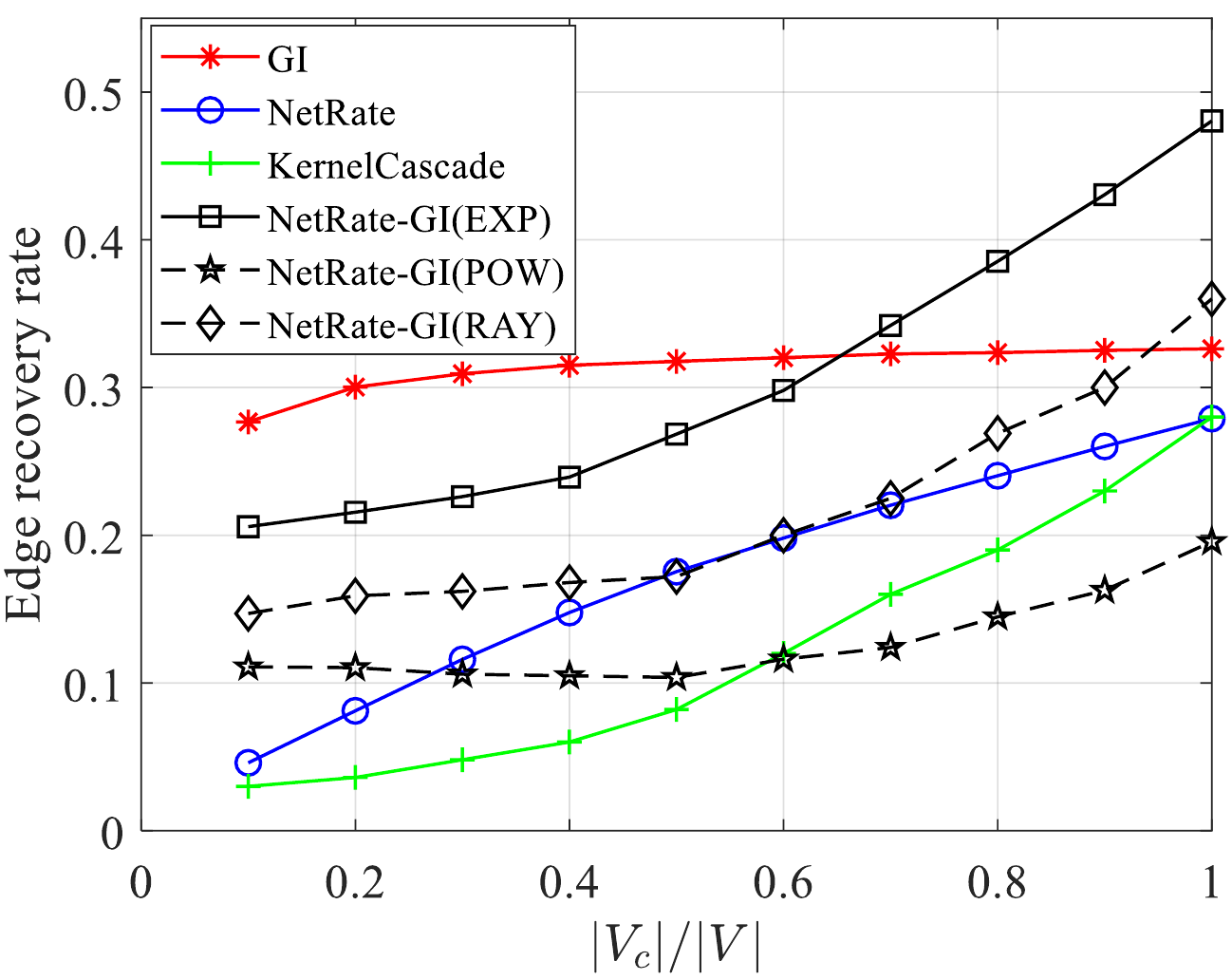}
			\centerline{(b)}
		\end{minipage}
		\caption{Performance comparison between NetRate, KernelCascade, GI and NetRate-GI on (a) Forest-fire network and (b) Email network with heterogeneous spreading distributions. The dashed curves show the performance of NetRate-GI if there is a distribution mismatch for comparison purpose.} \label{fig:hete_cases}
	\end{figure}

	From the solid curves in \cref{fig:hete_cases}, we observe that GI has the best performance when $|V_c|/|V|$ is small, while NetRate-GI has the best performance when $|V_c|/|V|$ is large. We also observe that in very dense networks like the Email network in \cref{fig:hete_cases}(b), GI performs the best over a large range of $|V_c|/|V|$ values. This is because NetRate was not able to accurately estimate the edge propagation time moments as well as the likelihood scores, which led to errors in NetRate-GI. However, the good performance of GI comes at the price of knowing the average mean and second order moment of the edge propagation times \emph{a priori}. We also test the effect of distribution mismatch on NetRate-GI. The results are shown in the same \cref{fig:hete_cases} by dashed curves. \cref{fig:hete_cases}(a) shows that even though there is distribution mismatch, the combined approach NetRate-GI (RAY) can still outperform the rest. NetRate-GI (POW) performs worst; we believe that is because NetRate (POW) produces very poor likelihood scores according to \cref{fig:methods_4}.

 \subsection{Bimodal spreading distributions}\label{subsec:bimodal}
		
    \blue{We also study the performance of GI when the spreading distributions are bimodal. A mixture of two normal distributions with equal standard deviations is bimodal if their means differ by at least twice the common standard deviation \cite{Mark2002}. In our simulations, we generate the propagation time along each edge from the normal distribution $\calN(\mu,\sigma^2)$ and the bimodal distributions $0.5\calN(\mu_1,\sigma^2) + 0.5\calN(\mu_2,\sigma^2)$, for different values of $\mu_1$ and $\mu_2$, where $|\mu_1-\mu_2|\geq2\sigma$. Let $|V_c|/|V|=0.5$, the results are shown in \cref{tab:bimodal}. We see that in the case of bimodal spreading distributions, the performance of GI is worse when $|\mu_1-\mu_2|$ is larger or when the network is denser.} 
   	\begin{table}[!htb]
   	\caption{Edge recovery rate comparison of three spreading distributions.}
   	\centering  
   	\begin{tabular}{lC{2cm}C{2cm}C{2cm}C{2cm}C{2cm}}  
   		\hline
   		Distribution & E-R tree & E-R graph ($\text{deg}_{\text{ave}}=4$) & E-R graph ($\text{deg}_{\text{ave}}=8$) & Forest-fire network & Email network\\ 
        \hline
   		$\calN(5,1)$ &0.98 &0.97 &0.59 &0.48 &0.32\\
     
   		$0.5\calN(3,1)+0.5\calN(7,1)$ &0.96 &0.68 &0.48 &0.42 &0.17\\   

   		$0.5\calN(1,1)+0.5\calN(9,1)$ &0.74 &0.42 &0.28 &0.11 &0.08\\   
   		\hline
   	\end{tabular}
   	\label{tab:bimodal}
    \end{table}    

\subsection{Real dataset}\label{subsec:real_world}

Finally, we use the MemeTracker dataset \cite{Gom2011} from SNAP to compare GI, NetRate, KernelCascade and NetRate-GI. MemeTracker builds maps of the daily news cycle by analyzing around 9 million news stories and blog posts per day from 1 million online sources. We use hyperlinks between articles and posts to represent the flow of information from one site to other sites. A site publishes a new post and puts hyperlinks to related posts published by some other sites at earlier times. At a later time, this site's post can also be cited by newer sites. This procedure is then repeated and we are able to obtain a collection of timestamped hyperlinks between different sites (in blog posts) that refer to the same or closely related pieces of information. This collection of timestamps is recorded as a cascade. 
		
\blue{We divide the dataset into two parts. From the first part, we extract a sub-network with top 500 sites and 2457 edges, which contains blog posts in two months. This is used as the historical data to estimate the moments of edge propagation times required by our algorithm. If one site re-posted a blog published by another site, we connect an edge between these two sites and obtain the propagation time according to the timestamp information. From this sub-network, we estimate the mean and second order moment of the edge propagation time as 0.46 second and 1.29 second squared respectively. We then use the second part of the dataset to extract 2400 cascades from 1,272,031 posts in a month to test different algorithms (notice that for each cascade, only a subset of vertices are timestamped).} 

For NetRate and NetRate-GI, we assume an exponential diffusion model. The comparison results are shown in \cref{tab:real_data}. By comparison, NetRate-GI has the best performance when the number of cascades is at least 1600, with GI in second place. GI however has the best performance when the number of cascades is small. 
    
\blue{For other applications, the historical data we need to estimate the edge propagation time moments is similar to the first part of the dataset described above. The historical dataset's size can be small and may not even come from the same source. For example, while trying to infer the topology of a Facebook sub-network, we can use data from another known Facebook sub-network or another social network like Twitter to estimate the moments. We see from results in \cref{subsec:Heterogeneous,subsec:bimodal} that GI is relatively robust to errors in the moment estimation.}
		
 	\begin{table}[!htbp]
 	    \caption{Comparison of edge recovery rate on the MemeTracker dataset.}
	 	\centering  
	 	\begin{tabular}{lcccc}  
	 		\hline
	 		Number of cascades &GI &NetRate (EXP) & KernelCascade &NetRate-GI (EXP)\\
	 		\hline
	 		800 &0.32 &0.24 &0.14 & 0.27  \\       
	 		1600 &0.53 &0.46 &0.35 & 0.57 \\ 
	 		2400 &0.69 &0.63 &0.67 & 0.76\\  
	 		\hline
	 	\end{tabular}
	 	\label{tab:real_data}
	 \end{table}

	\section{Conclusion} \label{sec:con}
	
	In this paper, we have developed a theory and method for graph topology inference using information cascades and knowledge of some moments of the diffusion distribution across each edge, without needing to know the distribution itself. In the case of tree networks, we provided a necessary condition for perfect reconstruction, and used the concept of redundant vertices to propose an iterative tree inference algorithm. Simulations demonstrate that our method outperforms the tree reconstruction algorithm in \cite{Abr2013}. We have also provided some theoretical insights into how the moments of the propagation time between two vertices in a general graph behave, and extended our tree inference method heuristically to general graphs. Our simulation results suggest that our graph inference algorithm performs reasonably well, if the total number of cascades is not too small compared to the size of the network, and often outperforms the NetRate algorithm in \cite{Gom2011} and KernelCascade algorithm in \cite{Dunips}. Moreover, our method is suitable for time-critical applications owing to its low complexity. 
	
\appendices

\section{Proof of Theorem \ref{thm:lvca}}\label[appendix]{proof:thm:lvca}
	
		(a) Suppose that $u,v$ are connected by an edge $e$ in $E$. For each $v_i$, let $P_u$ be a geodesic connecting $v_i$ and $u$. Concatenating $P_u$ with $e$ gives a path (not necessarily simple) connecting $v_i$ and $v$, and therefore $d_{v_i}(v)\leq d_{v_i}(u)+1$. The same argument switching the roles of $u$ and $v$ gives $d_{v_i}(u)\leq d_{v_i}(v)+1$. Part (a) thus follows.
		
		(b) Let $u,v \in \conv(V')$ be two distinct vertices. By the definition of convex hull, we can find four vertices $v_1,v_2,v_3,v_4 \in V'$ (not necessarily distinct) such that $u \in [v_1,v_2]$ and $v \in [v_3,v_4]$. Because $G$ is a tree, there is a unique simple path $P$ connecting $[v_1,v_2]$ and $[v_3,v_4]$ if they are disjoint. If $[v_1,v_2]$ and $[v_3,v_4]$ have a non-empty intersection, then take $P$ to be any vertex in the intersection. Let $u' = P\cap [v_1,v_2]$ and $v' = P\cap [v_3,v_4]$. Without loss of generality, we assume that $u \in [v_1,u']$ and $v \in [v',v_4]$. Therefore, $[v_1,v_4] = [v_1,u']\cup [u',v']\cup [v',v_4]$, and hence $u,v\in [v_1,v_4]$. Consequently, $d_{v_1}(u)\neq d_{v_1}(v)$ and $d_{v_4}(u) \neq d_{v_4}(v)$. By definition, $V'$ separates $\conv(V')$.  
		
		(c) By part (a), it suffices to show that $E'\subset E$. Let $u$ and $v$ be two vertices connected by an edge in $E'$. This means that for all $i=1,\ldots,l$, we have $|d_{v_i}(u)-d_{v_i}(v)|\leq 1$. Suppose on the contrary that $u$ and $v$ are not connected by an edge in $E$. Because $V'$ separates $G$, we have that each connected component of $G \setminus \conv(V')$ is a simple path; for otherwise, there will be two distinct vertices of $G \setminus \conv(V')$ having the same distance to all $v_i\in V'$.
		
		We first claim that both $u$ and $v$ are not in $\conv(V')$. If on the contrary, $u \in \conv(V')$, let $v'$ be the vertex in $\conv(V')$ closest to $v$ (called \emph{the projection of $v$ onto $\conv(V')$}). As in the proof of (b), we see that there are $v_1, v_2 \in V'$ such that $[v',u]\subset [v_1,v_2]$. Without loss of generality, assume that $u \in [v',v_1]$. Therefore $u \in [v,v_1]$ and $|d(v,v_1)-d(u,v_1)|\leq 1$ if and only if $d(u,v)=1$.
		
		If $u$ and $v$ are in the same component of $G\setminus \conv(V')$, then the condition $|d_{v_i}(u)-d_{v_i}(v)|\leq 1$ easily implies that $u$ and $v$ are connected by an edge (notice that each component of $G\setminus \conv(V')$ is a simple path if $V'$ separates $G$), which gives a contradiction.
		
		Next, assume that $u$ and $v$ are in different components of $G\setminus \conv(V')$. Let $u_1$ and $u_2$ be their respective closest vertex (projections) on $\conv(V')$. Notice that $u_1,u_2 \in \partial B_G'\cup \partial \conv(V')$. According to the given condition, $d(u_1,u_2)>1$. Without loss of generality, we assume that $d(u,u_1)\geq d(u_2,v)$. Moreover, we have seen in the proof of (b) that we can choose $v_1,v_2 \in V'$ such that $[u_1,u_2]\subset [v_1,v_2]$ and $u_1\in [v_1,u_2]$. Therefore,
		\begin{equation*}
		\begin{split}
		d_{v_2}(u)-d_{v_2}(v) & =d(u_2,u)-d(u_2,v) \\ & = d(u_2,u_1)+d(u,u_1)-d(u_2,v)>1.
		\end{split}
		\end{equation*}
		This contradicts the fact that $u$ and $v$ are connected by an edge in $E'$.
		
		(d) Suppose on the contrary that $V'$ does not separate $G$. We can find two vertices $u_1$ and $u_2$ such that $d_{v_i}(u_1) = d_{v_i}(u_2), v_i \in V'$. Let $u_3$ be a vertex connected by an edge to either $u_1$ or $u_2$, say $u_1$. Then for each $v_i\in V'$, $|d_{v_i}(u_3) - d_{v_i}(u_2)| \leq |d_{v_i}(u_3) - d_{v_i}(u_1)|+|d_{v_i}(u_1) - d_{v_i}(u_2)|\leq 1$. Therefore, because $E=E'$, $u_3$ and $u_2$ are also connected by an edge. This contradicts the assumption that $G$ does not contain any triangle. The proof is now complete.

	\begin{figure}[!tb]
		\centering 
		\begin{minipage}[b]{0.5\linewidth}
			\centering
			\includegraphics[scale=0.6]{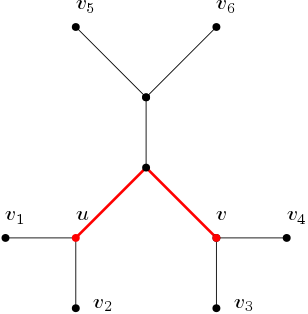}
		\end{minipage}%
		\begin{minipage}[b]{0.5\linewidth}
			\centering
			\includegraphics[scale=0.7]{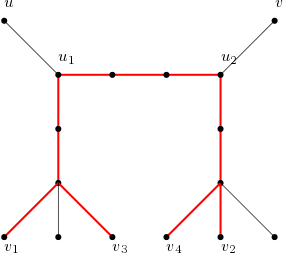}
		\end{minipage}
		\caption{The example on the left illustrates the proof of \cref{thm:lvca}(b). The red path is $P$ given in the proof; moreover, $u'=u$ and $v'=v$. The example on the right illustrates the proof of \cref{thm:lvca} (c), when $u,v$ are in different compoents of $G\setminus \conv(V')$. The red edges form the convex hull of $V'=\{v_1,\ldots,v_4\}$. Because $u$ and $v$ are not connected by an edge, in the proof, we find $v_1$ (or $v_2$) such that $|d(v_1,u)-d(v_1,v)|>1$.}
		\label{fig:3plus6}
	\end{figure}	
	
%

%

\section{Supplementary discussions to Section \ref{sec:sep}} \label[appendix]{sup}

In this appendix, we provide further insights into the results in \cref{sec:sep,sec:re}. We start with a technical definition. 

\begin{Definition_A}
In the graph $G$, if $v$ is a leaf and the unique neighbor of $v$ is ordinary, then we say that $v$ is a \emph{long leaf}. The set of long leaves is denoted by $L_{\partial B_G}$. 
\end{Definition_A}

As an example, in \cref{fig:1}, $v_1, v_2$, and $v_3$ are long leaves, and $L_{\partial B_G}=\{u_1\}$ (note that $u_2\notin L_{\partial B_G}$). Given a graph $G$, the size of $L_{\partial G}$ and $L_{\partial B_G}$ can be easily computed. 

\begin{figure}[!htb]
\centering
\begin{minipage}[b]{.5\linewidth}
\centering
\centerline{\includegraphics[scale=0.6]{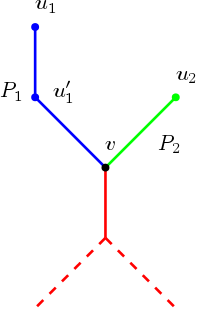}}
\centerline{(a)}
\end{minipage}%
\begin{minipage}[b]{.5\linewidth}
\centering
\centerline{\includegraphics[scale=0.625]{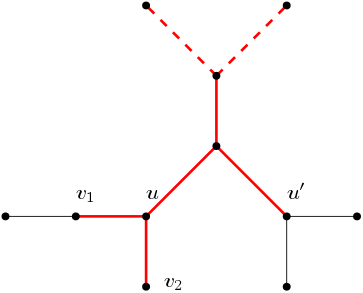}}
\centerline{(b)}
\end{minipage}
\caption{The figure (a) illustrates the proof of \cref{coro:lvst}. The red edges form $\conv(V')$. The two paths $P_1$ and $P_2$ in the proof are the blue and green paths, respectively. The figure (b) illustrates the proof of \cref{coro:sgiat}. The red edges form $\conv(V')$. According to the construction, $L_u=\{v_1,v_2\}$ and $L_{u'}=\{u'\}$. We can remove $v_1$ from $V'$.}
\label{fig:12plus14}
\end{figure}

\begin{Corollary_A} \label{coro:lvst}
Suppose that $G=(V,E)$ is a tree. Given $V' = \{v_1,\ldots,v_l\} \subset V$ and $\{d_{v_i}(\cdot): v_i\in V'\}$, suppose that the following uniqueness property holds: for any tree $G'=(V,E')$ spanning $V$ with the associated distance function $d'(\cdot,\cdot)$ defined on $E'$, if $d'(v_i,u) = d_{v_i}(u)$ for any $v_i\in V'$ and $u\in V$, then $G'=G$. Then, $l\geq |L_{\partial G}|-|L_{\partial B_G}|$. In particular, if $l < |L_{\partial G}|-|L_{\partial B_G}|$, then $\psi(V')<1.$
\end{Corollary_A}
\begin{IEEEproof}
Suppose on the contrary that $l< |L_{\partial G}|-|L_{\partial B_G}|$. By the pigeon-hole principle and \cref{thm:lvca}(b), we can always find two simple paths $P_1$ and $P_2$ in $G\setminus \conv(V')$ that are disjoint except where they intersect at a vertex $v\in B_G$, and the length of at least one of them, say $P_1$, is larger than $1$ (see \cref{fig:12plus14}(a)).

Therefore, we can find $u_1 \in P_1$ such that $d(u_1,v)=2$, and $u_2 \in P_2$ such that $d(u_2,v)=1$. Moreover, let $u_1'$ be the vertex between $u_1$ and $v$. Because both $P_1$ and $P_2$ are not in $\conv(V')$, it is impossible to determine if $u_1$ is connected to $u_1'$ or $u_2$ based only on the information $\{d_{v_i}(\cdot) : v_i \in V'\}$, which contradicts the uniqueness property. The claim therefore holds. From \cref{coro:teg}\ref{it:tega}, $\psi(V')=1$ implies the uniqueness property, and the last statement of the corollary follows from contra-positiveness. 
\end{IEEEproof}

\Cref{coro:lvst} provides a necessary condition for the minimum size of a separating set for a tree. 

\section{Proofs of results in Section \ref{sec:re}}\label[appendix]{proof:sec:re}

We prove results stated in \cref{sec:re}.

\begin{IEEEproof}[Proof of \cref{lem:avvi}]
(a) If $|d_{v_i}(u)-d_{v_i}(v)|>1,$ then $u,v$ are not connected to each other when we use $V'$. Therefore, $v_i$ is redundant if and only if they are not connected to each when we use $V'\backslash \{v_i\}$; or equivalently there is some $v_j \in V'\backslash \{v_i\}$ such that $|d_{v_j}(u)-d_{v_j}(v)|>1.$

(b) This follows immediately from the criterion given in \ref{it:avvi}.

(c) According to the definition, $E'$ constructed from $V'\backslash \{v_1\}$ and $V'\backslash \{v_2\}$ are both the same as that constructed from $V'$.
\end{IEEEproof}

\begin{IEEEproof}[Proof of \cref{prop:sivi}]

\begin{figure}[!htb] 
\centering
\includegraphics[scale=0.6]{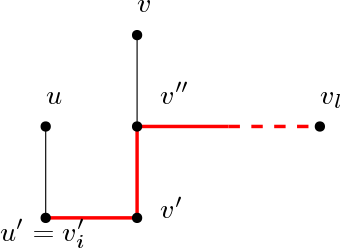}
\caption{This figure illustrates the proof of \cref{prop:sivi} (b) when $d(u',v')\leq 1$. The red edges form $\conv(V')$. The other nodes follow the same notation as given in the proof; $v_l$ is the node of $V'$ such that $|d(v_l,u)-d(v_l,v)|>1$.} \label{fig:13}
\end{figure}

Suppose that we are given $u,v\in V$ such that $|d_{v_i}(u)-d_{v_i}(v)|>1$. Let $u'$ and $v'$ be their respective closest points in $\conv(V'\setminus\{v_i\})$. We have seen that we can always find $v_j, v_k \in V'\setminus\{v_i\}$ such that $[u',v']\subset [v_j,v_k]$. Without loss of generality, we assume that $u'\in [v_j, v']$. We notice that $[u,u']\cap \conv(V'\setminus\{v_i\}) = u'$ because $u'$ is the closest point; similarly, $[v, v']\cap \conv(V'\setminus\{v_i\}) = v'$. 

Suppose that $d(u',v')>1$ and without loss of generality that $d(u,u')\leq d(v,v')$. Therefore, we find $|d_{v_j}(v)-d_{v_j}(u)|\geq d(u',v')>1$. We are done in view of \cref{lem:avvi} (a). 

For the remaining case where $d(u',v')\leq 1$, we treat (a) and (b) separately.

(a) Suppose that $d(u',v')\leq 1$ and $v_i=v_i'$. Without loss of generality, we can further assume in this case that $u'\in [v_i,v']$ (notice that this requires that $d(u',v')\leq 1$). In this case, we see that 
\[
|d_{v_j}(v)-d_{v_j}(u)|=|d(u',u)-d(u',v)|=|d_{v_i}(v)-d_{v_i}(u)|>1.
\]

(b) If both $u'$ and $v'$ are different from $v_i'$, then the same argument as above does the job because, in this case, $|d_{v_i'}(v)-d_{v_i'}(u)|=|d_{v_i}(v)-d_{v_i}(u)|$. 
Suppose that $u'= v_i'$ (see \cref{fig:13}). By the definition of convex hull and the choice of $u', v'$, we have $[u',v']\subset [u,v]$. Let $v''$ be the neighbor of $v'$ on $[v,v']$; it is immediate that $d(v'',u)=d(v'',v_i')\leq 2$. Notice that $P$ contains ordinary vertices. Therefore, the assumption asserts that there is a $v_l\in V'\setminus\{v_i\}$ such that $d(v_l,u')\geq 2$ and $v''\in [v',v_l]$. 

If $|d(u,u')-d(v,v')|>2$, then either use $v_k$ or $v_j$ to satisfy \cref{lem:avvi} (1). Otherwise, $|d(v'',u)-d(v'',v)|>1$. It is easy to verify that 
\[
|d_{v_l}(u)-d_{v_l}(v)|\geq |d(v'',u)-d(v'',v)|>1.
\]
\end{IEEEproof}

\begin{IEEEproof}[Proof of \cref{coro:sgiat}]
Let $T$ be the convex hull of $V'$ and $L$ the leaves of $T$. For each $u \in \partial B_G'\cap T$, we define the set $L_u\subset L$ as follows: $$L_u = \{w \in [v,u]\cap L \mid v\in \partial G \text{ such that } [v,u]\cap B_G = \{u\}\}.$$ In other words, $L_u$ consists of the leaves of $T$ to whom $u$ is the closest among all the members of $B_G$. 

Suppose that $|L_u| = deg(u)-1$. If a neighbor $v$ of $u$ is not a leaf, then let $l_u$ be the (unique) node in $L_u$ such that $v\in [l_u,u]$; and otherwise, let $l_u$ be any node in $L_u$. Form $V''$ by removing these $l_u$ from $V'$ (see \cref{fig:12plus14}(b)). It is clear that $|V''|\leq |\partial G| - |\partial B_G'|$. Moreover, each other vertex is redundant by \cref{prop:sivi}(a) and (b). 
\end{IEEEproof}

\begin{IEEEproof}[Proof of \cref{eg:asy} \cref{it:iwc}]
Suppose that $u$ and $v$ are not connected by an edge using $v_1$, but are connected by an edge using $v_2$. This can only happen when $|d(u,v_2) - d(v,v_2)|=1$ and $|d(u,v_1)-d(v,v_1)|=3$. The same argument as in the proof of \cref{prop:sivi} proves the claim (use a neighbor of $v_2$ in place of $v''$ in the last two paragraphs; and see \cref{fig:15} for an illustration). 
\end{IEEEproof}

\section{Proofs of results in Section \ref{sec:graph}}\label[appendix]{proof:sec:graph}

\begin{IEEEproof}[Proof of \cref{thm:stt}]
We prove the theorem by gradually modifying the graph $G$ to a graph that is more symmetric so that $f_{u,v}$ can be explicitly written down. We start with the following elementary result.

\begin{Lemma_A} \label{lem:lmb}
Let $m,n, k$ be positive integers. Then for some $x\in(0,1)$, we have $$x^m+x^n-x^{m+n}>x^{1/k}.$$
\end{Lemma_A}
\begin{IEEEproof}
Set $h(x) = x^m+x^n-x^{m+n}-x^{1/k}.$ The first-order derivative is $$h'(x) = mx^{m-1}+nx^{n-1}-(m+n)x^{m+n-1}-1/kx^{1/k-1}.$$ We find immediately that $h'(1) = -1/k<0.$ As $h'(x)$ is continuous in a small neighborhood  containing $1$, we have $h'(x)<0$ in a small interval $(1-\epsilon, 1)$ for some $0<\epsilon <1/2$. Because $h(1)=0$, the mean value theorem allows us to conclude that $h(1-\epsilon)>0.$
\end{IEEEproof}

We now proceed with the proof of the theorem. It suffices to prove that $\bar{F}\neq \bar{F}_{u,v}.$ Suppose that the contrary is true. If $u$ and $v$ is connected by an edge, the mean propagation time will be strictly smaller than that of $f$ as we assume $f$ has infinite support (a more general result is given in \cref{lem:YX} below). For the rest of the proof, we assume $(u,v)\notin E$.

\begin{figure}[!htb] 
\centering
\includegraphics[scale=0.65]{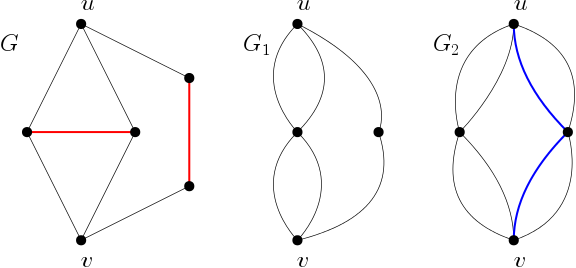}
\caption{This illustrates the proof of \cref{thm:stt}. In $G$, we shrink the two red edges to points, giving $G_1$. Two additional blue edges are added in $G_1$ to give $G_2$.} \label{fig:10}
\end{figure}

We first reduce the graph $G$ to a simpler graph (see \cref{fig:10} for illustration). Construct $G_1$ as follows: if both end points of an edge $e$ in $G$ are different from $u$ and $v$, then shrink $e$ to a single point (take note that this is different from removing $e$) in $G_1$. As $u,v$ are not connected by an edge in $G$, each path between $u$ and $v$ is of length $2$ in $G_1$. We should take note that multiple edges are allowed between two nodes in $G_1$ (and $G_2$ constructed below). The shrinking process reduces the time to travel from $u$ to $v$ and hence $\bar{F}_{u,v} \geq \bar{F}_1.$

Let $C$ be a connected component of $G_1\backslash \{u,v\}$. It is easy to see that $C$ is made up of a single node $v_C$, with $m_C$ paths connecting $v_C$ to $u$, and $n_C$ paths connecting $v_C$ to $v$. Denote the total number of connected components of $G_1\backslash \{u,v\}$ by $k$. Define $m=\max_C\{m_C\}$ and $n=\max_C\{n_C\}$. Construct $G_2$ as follows: for each $C$, we add $m-m_C$ edges between $v_C$ and $u$, and add $n-n_C$ edges between $v_C$ and $v$. As we add additional edges between nodes without changing the rest of the graph, we have $\bar{F}_1\geq \bar{F}_2.$ The number of connected components of $G_2\backslash \{u,v\}$ is still $k$.

Let $f^{[n]}$ be the first-order derivative of $1-\bar{F}^n$. As $\bar{F}(x)=1$ if $x\leq 0$, for $t>0$, we have the following
\begin{align*}
& \bar{F}_{u,v}(t) \geq \bar{F}_1(t) \geq \bar{F}_2(t) \\
& = \left(\int_0^{\infty}\bar{F}^m(t-x)f^{[n]}(x) d x\right)^{k}\\
& = \left(\int_0^{t}\bar{F}^m(t-x)f^{[n]}(x) d x + \int_t^{\infty}\bar{F}^m(t-x)f^{[n]}(x) d x\right)^{k} \\
& = \left(\int_0^{t}\bar{F}^m(t-x)f^{[n]}(x)d x + \int_t^{\infty}f^{[n]}(x)d x\right)^{k} \\
& \geq \left(\bar{F}^m(t)\int_0^{t}f^{[n]}(x)d x + \bar{F}^n(t)\right)^{k} \\
& = \left(\bar{F}^m(t)(1-\bar{F}^n(t))+\bar{F}^n(t)\right)^k \\
& = \left(\bar{F}^m(t)+\bar{F}^n(t)-\bar{F}^{m+n}(t)\right)^k.
\end{align*}
Because $\bar{F}(t)$ is continuous, $\bar{F}(0)=1$ and $\lim_{t\to \infty}\bar{F}(t)=0$. By the intermediate value theorem, $(0,1]$ is contained in the image of $\bar{F}$. Therefore, from \cref{lem:lmb}, we obtain a contradiction and the theorem is proved.  
\end{IEEEproof}

\begin{IEEEproof}[Proof of \cref{lem:YX}]
We have
\begin{align*}
\E[Y_{u,v}^k] - \E[X_{u,v}^k] 
& = \int_0^{\infty}\int_0^{\infty}(y^k-\min(x,y)^k)h(y)f(x)dydx \\
& =  \int_0^{\infty}\int_{x}^{\infty}(y^k-x^k)h(y)f(x)dydx \\
& \geq \int_0^{\infty}\int_{(x^k+\epsilon_1)^{1/k}}^{\infty}(y^k-x^k)h(y)f(x)dydx \\
& \geq \epsilon_1\int_0^{\infty}\int_{(x^k+\epsilon_1)^{1/k}}^{\infty}h(y)f(x)dydx \\
& = \epsilon_1 \int_{\epsilon_1^{1/k}}^{\infty}\int_{0}^{(y^k-\epsilon_1)^{1/k}}f(x)h(y)dxdy \\ 
& = \epsilon_1\int_{\epsilon_1^{1/k}}^{\infty}F((y^k-\epsilon_1)^{1/k})h(y)dy \\
& \geq \epsilon_1 F((\epsilon_0-\epsilon_1)^{1/k}) \int_{\epsilon_1^{1/k}}^{\infty} h(y)dy \\
& = \epsilon_1 F((\epsilon_0-\epsilon_1)^{1/k})\bar{H}(\epsilon_0^{1/k}),
\end{align*}
where the interchange of integration in the third equality follows from Tonelli's theorem. The lemma is proved. 
\end{IEEEproof}

\begin{IEEEproof}[Proof of \cref{lem:lgb}]
For any vertices $a, b$, let $P_{a,b}$ be the path associated with $X_{a,b}$. The concatenation of $P_{u,w}$ and $P_{u,v}$ is a path from $w$ to $v$ with possibly some edges repeated. We therefore have almost surely, $$X_{w,u}+X_{u,v}\geq X_{w,v} \Rightarrow X_{w,v}-X_{w,u}\leq X_{u,v}.$$ Similarly, $X_{w,u}-X_{w,v} \leq X_{u,v}$ almost surely. We then obtain
\begin{align*}
|\E[X_{w,u}]-\E[X_{w,v}]|^k \leq \E[|X_{w,u}-X_{w,v}|^k] \leq \E[X_{u,v}^k],
\end{align*}
where the first inequality follows from Jensen's inequality, and the lemma is proved.
\end{IEEEproof}

\begin{IEEEproof}[Proof of \cref{lem:sxa}]
Let $p$ and $h$ be the pdf of $P$ and $H$ respectively. It is easy to show that the pdf of the distribution associated with $Z$ is $p(1-H)+h(1-P).$ Therefore, we have to show that for each $\epsilon>0$, $$\int_0^{\infty}|p(x)(1-H(x))+h(x)(1-P(x))-h(x)|d x \leq 2(\bar{H}(\epsilon)+P(\epsilon)).$$ To see this, we first apply the triangle inequality:
\begin{align}
& \int_0^{\infty}|p(x)(1-H(x))+h(x)(1-P(x))-h(x)| d x \nonumber\\
\leq & \int_0^{\infty}h(x)P(x)dx + \int_0^{\infty}p(x)(1-H(x)) d x. \label{ineq:Z}
\end{align}
The two integrals on the right hand side of \eqref{ineq:Z} can be bounded separately as:
\begin{equation*}
\begin{aligned}
\int_0^{\infty}h(x)P(x)dx & = \int_0^{\epsilon}h(x)P(x)dx + \int_{\epsilon}^{\infty}h(x)P(x)dx \\
& \leq P(\epsilon)\int_0^{\epsilon}h(x)dx + \int_{\epsilon}^{\infty}h(x)dx\\
& \leq P(\epsilon) + \bar{H}(\epsilon),
\end{aligned}
\end{equation*}
\begin{equation*}
\begin{aligned}
\int_0^{\infty}p(x)(1-H(x))dx & = \int_0^{\epsilon}p(x)(1-H(x))dx + \int_{\epsilon}^{\infty}p(x)(1-H(x))dx \\
& \leq \int_0^{\epsilon}p(x) + \bar{H}(\epsilon)\int_{\epsilon}^{\infty}p(x)dx \\
&\leq P(\epsilon) + \bar{H}(\epsilon).
\end{aligned}
\end{equation*}
The result follows by adding up the two inequalities.
\end{IEEEproof}

\bibliography{IEEEabrv,SIS}{}
\bibliographystyle{IEEEtran}
\end{document}

%% file: preamble.tex
\usepackage[T1]{fontenc}
\usepackage{amsmath,amssymb,amsfonts,mathrsfs,bm}
\usepackage{amsthm}
\usepackage{array}

\usepackage[shortlabels]{enumitem}
\usepackage{graphicx}
\usepackage{url}
\usepackage{color}
\usepackage{algorithm,algorithmic}
\usepackage{multirow}
\usepackage[normalem]{ulem}
\usepackage{array}
\newcolumntype{L}[1]{>{\raggedright\let\newline\\\arraybackslash\hspace{0pt}}m{#1}}
\newcolumntype{C}[1]{>{\centering\let\newline\\\arraybackslash\hspace{0pt}}m{#1}}
\newcolumntype{R}[1]{>{\raggedleft\let\newline\\\arraybackslash\hspace{0pt}}m{#1}}
\usepackage{xparse}

\ifx\notloadhyperref\undefined
\ifx\loadbibentry\undefined
\usepackage[hidelinks]{hyperref} 
\else
\usepackage{bibentry}
\makeatletter\let\saved@bibitem\@bibitem\makeatother
\usepackage[hidelinks]{hyperref}
\makeatletter\let\@bibitem\saved@bibitem\makeatother
\fi
\else
\relax
\fi

\usepackage[capitalize]{cleveref}
\crefname{equation}{}{}
\Crefname{equation}{}{}
\crefname{claim}{claim}{claims}
\crefname{case}{case}{cases}
\crefname{occasion}{occasion}{occasions}
\crefname{line}{line}{lines}
\crefname{Theorem}{Theorem}{Theorems}
\crefname{Corollary}{Corollary}{Corollaries}
\crefname{Proposition}{Proposition}{Propositions}
\crefname{Lemma}{Lemma}{Lemmas}
\crefname{Definition}{Definition}{Definitions}
\crefname{Table}{Table}{Tables}
\crefname{Example}{Example}{Examples}
\crefname{Assumption}{Assumption}{Assumptions}
\crefname{Remark}{Remark}{Remarks}
\crefname{Theorem_A}{Theorem}{Theorems}
\crefname{Corollary_A}{Corollary}{Corollaries}
\crefname{Proposition_A}{Proposition}{Propositions}
\crefname{Lemma_A}{Lemma}{Lemmas}
\crefname{Definition_A}{Definition}{Definitions}

\ifx\useTheoremCounter\undefined
\newtheorem{Theorem}{Theorem}
\newtheorem{Corollary}{Corollary}
\newtheorem{Proposition}{Proposition}
\newtheorem{Lemma}{Lemma}
\else
\newtheorem{Theorem}{Theorem}
\newtheorem{Corollary}[Theorem]{Corollary}
\newtheorem{Proposition}[Theorem]{Proposition}
\newtheorem{Lemma}[Theorem]{Lemma}
\fi

\newtheorem{Example}{Example}
\newtheorem{Remark}{Remark}
\newtheorem{Assumption}{Assumption}

\newtheorem{Lemma_A}{Lemma}[section]
\newtheorem{Definition_A}{Definition}[section]
\newtheorem{Corollary_A}{Corollary}[section]

\theoremstyle{remark}

\newcommand{\Real}{\mathbb{R}}

\newcommand{\calC}{\mathcal{C}}

\newcommand{\calN}{\mathcal{N}}

\newcommand{\calT}{\mathcal{T}}

\DeclareSymbolFont{bsfletters}{OT1}{cmss}{bx}{n}
\DeclareSymbolFont{ssfletters}{OT1}{cmss}{m}{n}
\DeclareMathSymbol{\bsfGamma}{0}{bsfletters}{'000}
\DeclareMathSymbol{\ssfGamma}{0}{ssfletters}{'000}
\DeclareMathSymbol{\bsfDelta}{0}{bsfletters}{'001}
\DeclareMathSymbol{\ssfDelta}{0}{ssfletters}{'001}
\DeclareMathSymbol{\bsfTheta}{0}{bsfletters}{'002}
\DeclareMathSymbol{\ssfTheta}{0}{ssfletters}{'002}
\DeclareMathSymbol{\bsfLambda}{0}{bsfletters}{'003}
\DeclareMathSymbol{\ssfLambda}{0}{ssfletters}{'003}
\DeclareMathSymbol{\bsfXi}{0}{bsfletters}{'004}
\DeclareMathSymbol{\ssfXi}{0}{ssfletters}{'004}
\DeclareMathSymbol{\bsfPi}{0}{bsfletters}{'005}
\DeclareMathSymbol{\ssfPi}{0}{ssfletters}{'005}
\DeclareMathSymbol{\bsfSigma}{0}{bsfletters}{'006}
\DeclareMathSymbol{\ssfSigma}{0}{ssfletters}{'006}
\DeclareMathSymbol{\bsfUpsilon}{0}{bsfletters}{'007}
\DeclareMathSymbol{\ssfUpsilon}{0}{ssfletters}{'007}
\DeclareMathSymbol{\bsfPhi}{0}{bsfletters}{'010}
\DeclareMathSymbol{\ssfPhi}{0}{ssfletters}{'010}
\DeclareMathSymbol{\bsfPsi}{0}{bsfletters}{'011}
\DeclareMathSymbol{\ssfPsi}{0}{ssfletters}{'011}
\DeclareMathSymbol{\bsfOmega}{0}{bsfletters}{'012}
\DeclareMathSymbol{\ssfOmega}{0}{ssfletters}{'012}

\DeclareMathOperator*{\argmin}{arg\,min}

\DeclareMathOperator{\conv}{conv}

\newcommand{\qednew}{\nobreak \ifvmode \relax \else
      \ifdim\lastskip<1.5em \hskip-\lastskip
      \hskip1.5em plus0em minus0.5em \fi \nobreak
      \vrule height0.75em width0.5em depth0.25em\fi}

\newcommand{\ofrac}[1]{{\frac{1}{#1}}}

\newcommand{\tc}[1]{^{(#1)}}

\newcommand{\cond}[2]{\left. {#1}\, \middle| \, {#2} \right.}

\DeclareDocumentCommand \P { g d() g } {%
	\IfNoValueTF {#3} 
	{%
		\IfNoValueTF {#1} 
		{%
			\IfNoValueTF {#2}
			{%
				\mathbb{P}%
			}%
			{%
				\mathbb{P}\left({#2}\right)%
			}%
		}%
		{%
			\IfNoValueTF {#2}
			{%
				\mathbb{P}_{#1}%
			}%
			{%
				\mathbb{P}_{#1}\left({#2}\right)%
			}%
		}%
	}%
	{%
		\IfNoValueTF {#1} 
		{%
			\mathbb{P}\left(\cond{#2}{#3}\right)%
		}%
		{%
			\mathbb{P}_{#1}\left(\cond{#2}{#3}\right)%
		}%
	}%
}

\DeclareDocumentCommand \E { g o g } {%
	\IfNoValueTF {#3} 
	{%
		\IfNoValueTF {#1} 
		{%
			\IfNoValueTF {#2}
			{%
				\mathbb{E}%
			}%
			{%
				\mathbb{E}\left[{#2}\right]%
			}%
		}%
		{%
			\IfNoValueTF {#2}
			{%
				\mathbb{E}_{#1}%
			}%
			{%
				\mathbb{E}_{#1}\left[{#2}\right]%
			}%
		}%
	}%
	{%
		\IfNoValueTF {#1} 
		{%
			\mathbb{E}\left[\cond{#2}{#3}\right]%
		}%
		{%
			\mathbb{E}_{#1}\left[\cond{#2}{#3}\right]%
		}%
	}%
}

\definecolor{gray90}{gray}{0.9}

\newcommand{\blue}[1]{{{\color{black} #1}}}

\newcommand{\hide}[1]{}

\graphicspath{{./Figures/}}